\documentclass[12pt, leqno]{article}
\usepackage[toc,page]{appendix}
\usepackage{amsmath,setspace,multirow,lineno}
\usepackage[font=small,labelfont=bf]{caption}
\usepackage[top=1.5in, bottom=1.5in, left=1.2in, right=1.2in]{geometry}
\usepackage[round]{natbib}
\usepackage{color, soul, bbm}

\DeclareCaptionStyle{italic}[justification=centering]{labelfont={bf},textfont={it},labelsep=colon}
\usepackage{graphicx,psfrag,epsf,textcomp,epstopdf, amsthm}
\usepackage{enumerate, dsfont}
\usepackage{natbib,url}
\usepackage[table]{xcolor}
\usepackage{booktabs, subfig, bm, paralist, mathpazo, tikz, longtable, microtype}
\usepackage[linesnumbered,ruled,vlined,longend]{algorithm2e}
\usepackage{algpseudocode}

\usepackage[pdftex, colorlinks=true]{hyperref}
\definecolor{darkblue}{rgb}{0,0,.6}
\definecolor{skyblue}{RGB}{135,206,235}
\hypersetup{citecolor=darkblue,linkcolor=darkblue,urlcolor=darkblue}
\definecolor{DarkRed}{rgb}{.7,0,.4}
\definecolor{ao(english)}{rgb}{0.0, 0.5, 0.0}

\usepackage{comment,orcidlink}
\DeclareMathAlphabet\mathbfcal{OMS}{cmsy}{b}{n}

\newcommand{\blind}{0}

\newcommand{\X}{\mathcal{X}}

\graphicspath{{plots/}}
\DeclareMathOperator*{\argmin}{\arg\!\min}

\newsavebox\CBox

\renewcommand\X{\mathcal{X}}

\newtheorem{@definition}{\sc Definition}[section]

\usepackage{array}
\newcommand{\PreserveBackslash}[1]{\let\temp=\\#1\let\\=\temp}
\newcolumntype{C}[1]{>{\PreserveBackslash\centering}p{#1}}
\newcolumntype{R}[1]{>{\PreserveBackslash\raggedleft}p{#1}}
\newcolumntype{L}[1]{>{\PreserveBackslash\raggedright}p{#1}}

\begin{document}

\def\spacingset#1{\renewcommand{\baselinestretch}{#1}\small\normalsize} \spacingset{1}

\if0\blind
{
\title{\bf Conformal prediction for functional time series: Application to age-specific mortality rates}}
\author{
\normalsize Han Lin Shang \orcidlink{0000-0003-1769-6430} \\
\normalsize Department of Actuarial Studies and Business Analytics \\
\normalsize Macquarie University \\
}
\date{}
\maketitle
\fi

\if1\blind
{
\title{\bf Conformal prediction for functional time series: Application to age-specific mortality rates}
} \fi

\date{\vspace{-5ex}}
\maketitle
   
\begin{abstract}
In demographic literature, forecast uncertainty is often quantified with a statistical model. This model-based approach may potentially suffer from drawbacks, namely model misspecification, selection effect, and lack of finite-sample validity. We introduce a model-agnostic and distribution-free procedure, conformal prediction, for constructing prediction intervals for a functional time series. In the family of conformal prediction, split conformal prediction divides the data into training, validation, and test sets. Within the validation set, we can select optimal tuning parameters by calibrating the empirical coverage probabilities to match their nominal values. With the selected optimal tuning parameters, we then construct the prediction intervals using the same forecasting model for the holdout data in the testing set. Without sample splitting, sequential conformal prediction sequentially updates the predicted quantiles via an autoregressive process. Using Australian age- and sex-specific log mortality rates, we evaluate and compare the interval forecast accuracy, as measured by empirical coverage probability, coverage probability difference and mean interval score, between the two variants of conformal prediction.

\vspace{.1in}
\end{abstract}

\newpage
\spacingset{1.57}

\section{Introduction}\label{sec:1}

In many developed countries, increases in longevity and an ageing population have raised concerns about the sustainability of pensions, health, and aged care systems \citep[see, e.g.,][]{Coulmas07}. These concerns have led to a surge of interest among government policymakers and planners in accurately modeling and forecasting age-specific mortality rates. In the demographic literature on human mortality, three mortality instruments are widely studied: mortality rate, survival function, and life-table death counts. Although these three mortality instruments are complementary \citep[see, e.g.,][]{PHG01, DHW09}, they differ by the amount and range of constraints. Most of the literature has focused on developing new approaches to modeling and forecasting age-specific mortality rates \citep[see, e.g.,][for comprehensive reviews]{Booth06, BT08, BCB23}; it is also the mortality instrument that we study.

Alongside explanation, prediction is one of the most inferential tasks for  demographers since it forms the basis of many key aspects of government policies. In the current literature on forecasting age-specific mortality rates, a gap exists in quantifying forecast uncertainty. Such an uncertainty is often manifested \textit{probabilistically} through the construction of prediction intervals \citep[see, e.g.,][]{Stoto83, Keilman90, AS05}. As was emphasized by \cite{Chatfield93, Chatfield00}, it is essential to provide interval forecasts as well as point forecasts to 
\begin{inparaenum}[1)]
\item assess future uncertainty levels;
\item enable different strategies to be planned for a range of possible outcomes indicated by the interval forecasts;
\item compare forecasts from different methods more thoroughly and
\item explore different scenarios based on various assumptions.
\end{inparaenum}

There are at least four ways to quantify forecast uncertainties: 
\begin{inparaenum}[1)]
\item one can compute total variability and construct prediction intervals based on normality \citep[see, e.g.,][]{HU07}, this approach relies on a parametric distributional assumption, which may be misspecified;
\item as a nonparametric alternative, one can use residual bootstrap method as studied in \cite{HS09} or time-series bootstrap methods, such as block bootstrap \citep{PPS19} or sieve bootstrap \citep{Paparoditis18, PS23}, to generate pseudo sample paths, but this approach can be computationally demanding; 
\item instead of bootstrapping, one can deploy Bayesian paradigm \citep{HM17}, to generate pseudo sample paths, but this approach can be computationally demanding and is subject to model misspecification risk even asymptotically; 
\item as a computationally fast nonparametric approach, one can implement a variant of conformal prediction, such as conformal prediction in \cite{LRW15} and \cite{DFV22} in statistics, \cite{HM21, Hong23} and \cite{SH25} in actuarial science.
\end{inparaenum}

Traditional conformal prediction often requires exchangeable residuals, excluding time series. For example, \cite{LRW15} introduced conformal prediction for a univariate independent and identically distributed functional data set, while \cite{DFV22} extended it to a multivariate functional data set. However, some recent works address this issue, such as adaptive conformal inference of \cite{GC21} or sequential conformal inference of \cite{XX23}, developed for distribution-shift time series. It is a valid procedure for time series with general dependency. Our contribution is to introduce conformal prediction for modeling and forecasting \textit{age-specific} mortality rates, not weekly death counts in \cite{DS24} or total fertility rates in \cite{DSH+24}. Between the two variants of conformal prediction, we recommend the sequential conformal prediction, which does not require sample splitting to calibrate, since it can update the predictive quantiles when new data arrive.

The paper is structured as follows: In Section~\ref{sec:2}, we introduce a motivating data set, namely Australian age- and sex-specific mortality rates from 1921 to 2021. Using a functional time-series forecasting method of \cite{HU07} described in Section~\ref{sec:3.0}, we introduce conformal prediction by revisiting the split conformal prediction in Section~\ref{sec:3.1} and then introduce sequential conformal prediction in Section~\ref{sec:3.2}. In Section~\ref{sec:4.1}, we implement either an expanding- or a rolling-window forecast scheme described. As measured by the empirical coverage probability (ECP) and its related metrics in Section~\ref{sec:4.2}, we evaluate and compare the interval forecast accuracy in Section~\ref{sec:4.3}. The conclusion is presented in Section~\ref{sec:5}, along with some ideas on how the methodology can be further extended.

\section{Age-specific mortality rates}\label{sec:2}

We demonstrate the conformal prediction using Australian age- and sex-specific mortality rates from 1921 to 2021, sourced from \cite{HMD25}. The original dataset captures mortality rates for both sexes. For analytical refinement, age groups have been aggregated to span 0-99 in a single year of age, with the last age group 100+. This aggregation helps generate a comprehensive overview of mortality trends. By examining changes in mortality rates as a function of age and year, it becomes evident that mortality rates have undergone substantial variation over the years.

\begin{figure}[!htb]
\centering
\subfloat[Female raw log mortality rates]
{\includegraphics[width=8.65cm]{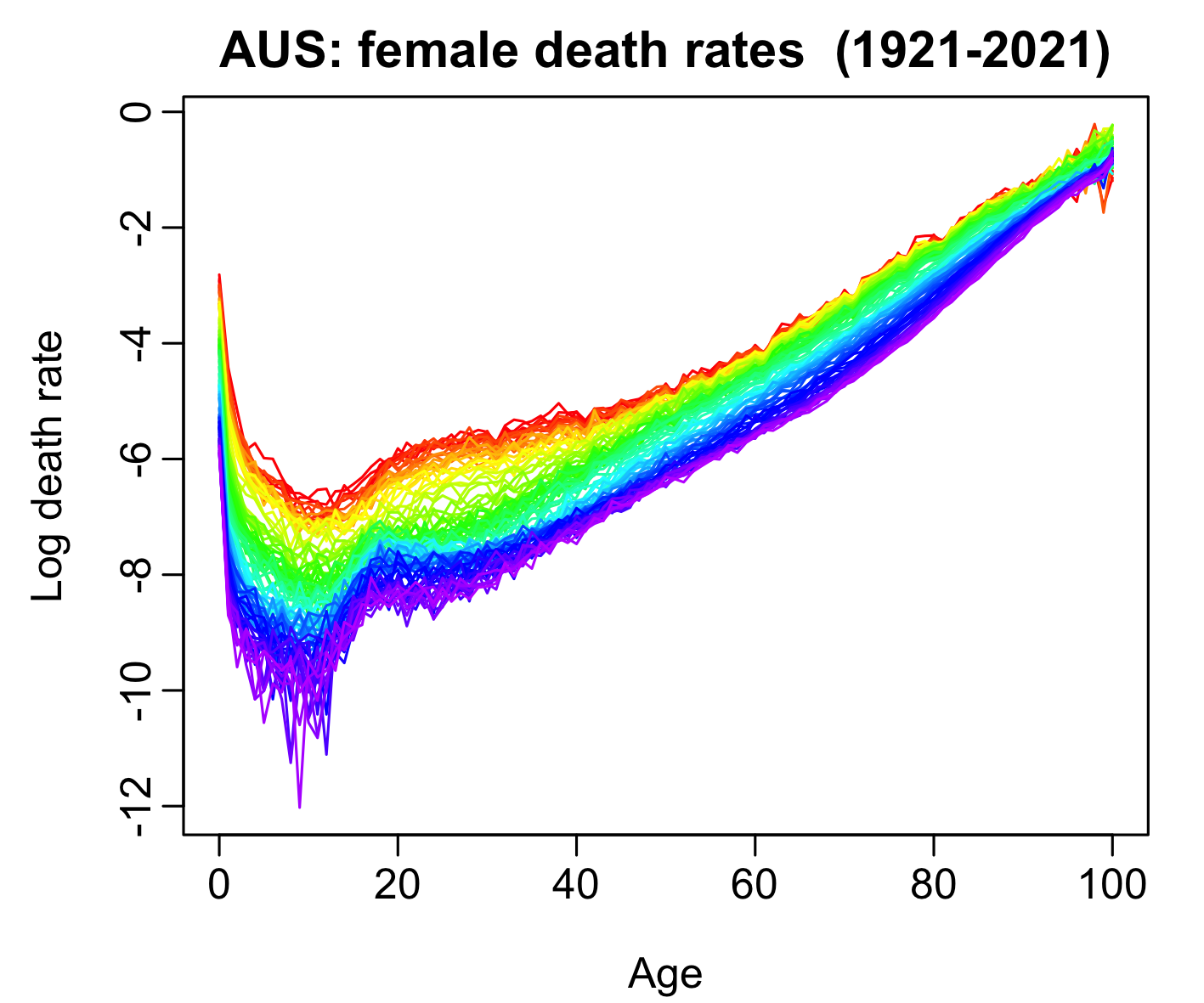}}
\quad
\subfloat[Male raw log mortality rates]
{\includegraphics[width=8.65cm]{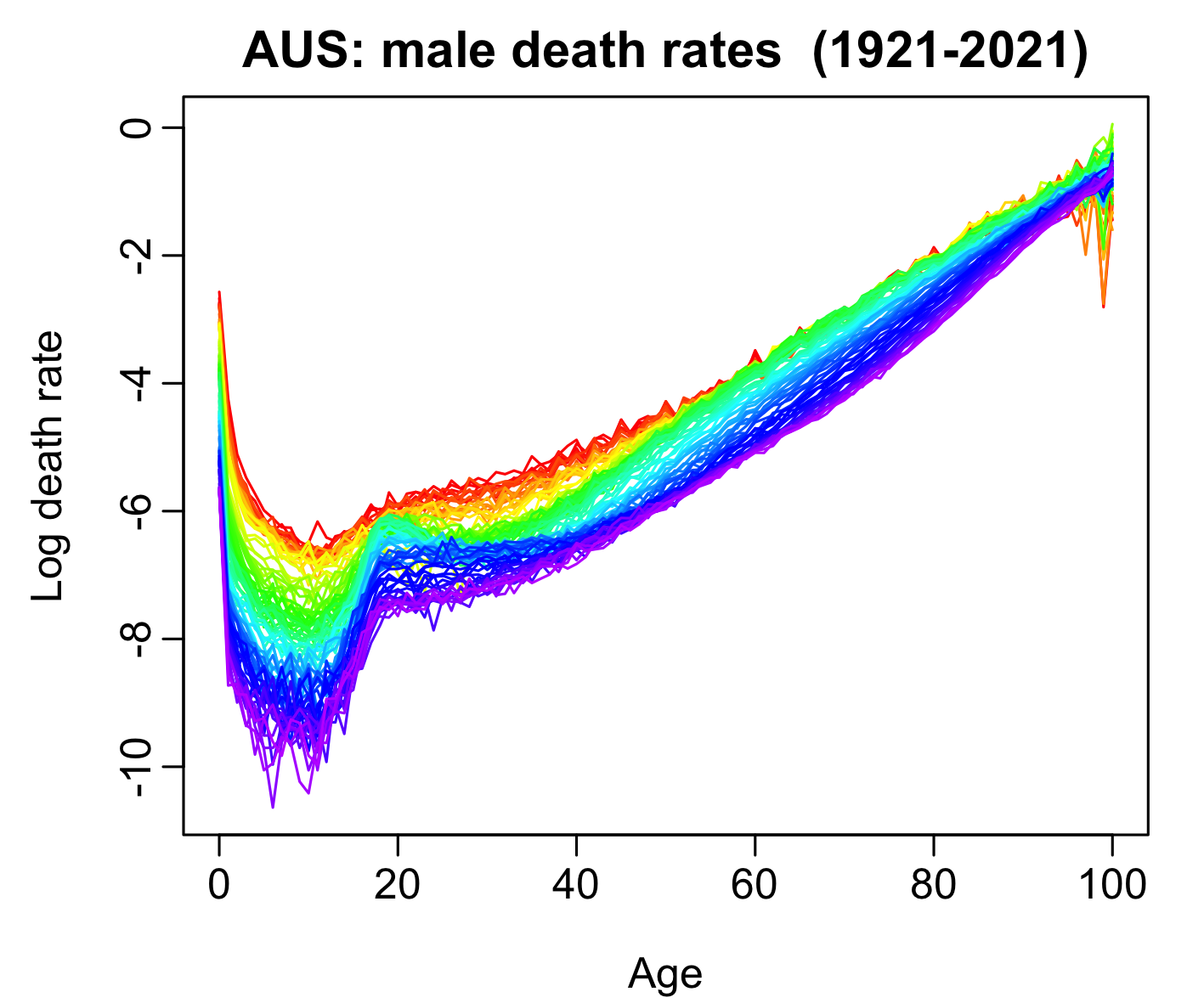}}
\caption{\small Rainbow plots of Australian age-specific female and male mortality rates from 1921 to 2021 by single year of age from 0 to 100 and the last age group of 100+. The aggregation helps generate a comprehensive overview of mortality trends. Data from the distant past are shown in red, while data from the most recent years are shown in purple.}\label{fig:1}
\end{figure}

To illustrate the historical trajectory, we show the age-specific mortality rates using a rainbow plot of \cite{HS10} in Figure~\ref{fig:1}. Log mortality rates from the distant past years are shown in red, while data from the most recent years are shown in purple. The figures show typical mortality curves for a developed country, with rapidly decreasing mortality rates in the early years of life, followed by an increase during the teenage years, a plateau in young adulthood, and then a steady increase from about age 30. In Australia, females have lower mortality rates than males at all ages.

\vspace{-.3in}

\section{Conformal prediction}\label{sec:3}

Introduced in the field of machine learning, the conformal prediction of \cite{SV08} and \cite{FZV23} is used to construct probabilistic forecasts calibrated on out-of-sample errors. Since its introduction in \cite{GVV98}, there have been many advances in the field, resulting in a variety of conformal prediction methods for various scientific applications, including time series forecasting \citep{ACT23}. In actuarial science, \cite{Hong23} and \cite{CLL+24} apply conformal prediction to credibility estimation and catastrophe bond pricing, respectively. In demography, for forecasting age distributions of death counts, \cite{SH25} recently focuses on split conformal prediction, in which prediction intervals are calibrated from a validation data set \citep{Cannon18} and assessed on a test data set.

\subsection{Functional time-series forecasting method}\label{sec:3.0}

Let $(\X_t: t\in \mathbb{Z})$ be an arbitrary functional time series, such as age-specific mortality rates in the natural logarithmic scale. It is assumed that the observations $\X_t$ are elements of the Hilbert space $\mathcal{H} = L^2(\mathcal{I})$ equipped with the inner product. Each function is a square-integrable function satisfying $\|\X_t\|^2 = \int_{\mathcal{I}} \X_t^2(u)du<\infty$. $\X(u)$ has a finite mean curve $\mu(u) = \text{E}[\X(u)]$ and a non-negative definite covariance function, given by
\begin{equation}
c_{\X}(u,v) = \text{cov}[\X(u),\X(v)] = \text{E}\{[\X(u) - \mu(u)][\X(v) - \mu(v)]\}\label{eq:1}
\end{equation}
for all $u, v \in \mathcal{I}$. The covariance function $c_{\X}(u,v)$ in~\eqref{eq:1} allows the covariance operator of $\X$, denoted by $K$, to be defined as
\begin{equation*}
K(\phi)(u) = \int_{\mathcal{I}} c_{\X}(u,v)\phi(v)dv.
\end{equation*}
Via the Mercer's lemma, there is an orthonormal sequence $(\phi_k)$ of continuous functions in $\mathcal{L}^2(\mathcal{I})$ and a non-increasing sequence $\lambda_k$ of positive numbers, such that
\begin{equation*}
c_{\X}(u,v) = \sum^{\infty}_{k=1}\lambda_k\phi_k(u)\phi_k(v),
\end{equation*}
where $\lambda_k$ is also known as the $k$\textsuperscript{th} eigenvalue. By the separability of Hilbert spaces, the Karhunen-Lo\`{e}ve expansion of a stochastic process $\X(u)$ can be expressed as
\begin{equation}
\X(u) = \mu(u) + \sum_{k=1}^{\infty}\beta_k\phi_k(u)	= \mu(u) + \sum_{k=1}^K\beta_k\phi_k(u)+e(u),\label{eq:2}
\end{equation}
where principal component scores $(\beta_1,\dots,\beta_K)$ are given by the projection of $[\X(u) - \mu(u)]$ in the direction of the $k$\textsuperscript{th} eigenfunction $\phi_k$, i.e., $\beta_k=\langle [\X(u) - \mu(u)], \phi_k(u)\rangle$; $e(u)$ denotes the model truncation error function with a mean of zero and a finite variance; and $K<n$ is the number of retained components. Equation~\eqref{eq:2} facilitates dimension reduction, as the first $K$ terms often provide a good approximation to the infinite sums, and thus the information inherited in $\bm{\X}(u)=\{\X_1(u),\dots,\X_m(u)\}$ can be adequately summarized by the $K$-dimensional vector $(\bm{\beta}_1, \bm{\beta}_2,\dots,\bm{\beta}_K)$. For theoretical, methodological, and applied aspects of functional principal component analysis, consult a number of survey articles by \cite{Hall11}, \cite{Shang14}, \cite{WCM16}, and \cite{RGS+17}.

To select the optimal number of components, we consider an eigenvalue ratio (EVR) criterion of \cite{LRS20} to determine $K$. Such an estimator is obtained simply by minimizing the ratio of two adjacent eigenvalues arranged in descending order. 
\begin{equation*}
K = \argmin_{1\leq k\leq k_{\max}}\left\{\frac{\lambda_{k+1}}{\lambda_k}\times \mathds{1}\{\lambda_k>\tau\}+\mathds{1}\{\lambda_k<\tau\}\right\},
\end{equation*}
where $\mathds{1}\{\cdot\}$ represents the binary indicator function. Customarily, $\tau=10^{-3}$ and $k_{\max}$ can be set as the number of curves $m$. For comparison, we set $K=6$ as in \cite{HBY13}.

Conditioning on the observed data $\bm{\X}(u)$ and the estimated functional principal components $\bm{B} = \{\widehat{\phi}_1(u),\dots,\widehat{\phi}_K(u)\}$, the $h$-step-ahead point forecast of $\X_{m+h}(u)$ can be expressed as
\begin{equation*}
\widehat{\X}_{m+h|m}(u) = \text{E}[\X_{m+h}(u)|\bm{\X}(u),\bm{B}]  =\widehat{\mu}(u) + \sum^K_{k=1}\widehat{\beta}_{m+h|m,k}\widehat{\phi}_k(u),
\end{equation*}
where $m$ represents a jump-off year, and $\widehat{\beta}_{m+h|m,k}$ denotes time-series forecasts of $k$\textsuperscript{th} principal component scores. Forecasts for these scores can be obtained using a univariate time-series forecasting method, such as exponential smoothing (ETS). Their model specifications can be automatically selected via the automatic algorithm of \cite{HK08}, which chooses the optimal orders based on an information criterion, such as the corrected Akaike information criterion \citep{HT93}.

\subsection{Split conformal prediction}\label{sec:3.1}

We divide the data sample consisting of 101 years from 1921 to 2021 into training, validation, and test sets, with split proportions of 60\%, 20\%, and 20\%. Using the training sample, we implement an expanding-window or a rolling-window forecast scheme to obtain the $h$-step-ahead forecasts in the validation set for $h=1,2,\dots,20$. The expanding-window scheme iteratively increases its training sample size, while the rolling-window scheme fixes a chosen sample size. Both forecast schemes allow us to assess how a forecasting method performs on relatively short to medium horizons. 

We have different numbers of curves in the validation set for each forecast horizon. For example, when $h=1$, we have 21 years to evaluate the residual functions; when $h=20$, we have two years to evaluate the residual functions, i.e., differences between the samples in the validation set and their forecasts. From these residual functions, we compute their pointwise summary statistics $\gamma_{h}(u)$, such as standard deviation, interquartile range (IQR), and mean absolute deviation (MAD), which are non-negative values of 101 ages. Compared to the standard deviation, IQR and MAD are robust against outliers \citep{LLK+13}. In the case of the quantile, we first compute the absolute values of the residuals and then take the $100\times (1-\alpha)\%$ quantile of the empirical distribution function. 

Forecast errors are denoted by $\widehat{\epsilon}_m(u) = \X_{m}(u) - \widehat{\X}_m(u)$, $m=1, 2,\dots,M$, and $M$ denotes the number of years of residual functions in the validation set. For a level of significance $\alpha$, our aim is to determine $(\underline{\xi}_{\alpha, h}, \overline{\xi}_{\alpha, h})$ such that $(1-\alpha)\times 100\%$ of the residuals satisfy
\begin{equation*}
-\underline{\xi}_{\alpha, h} \gamma_{h}(u)\leq \widehat{\epsilon}_m(u) < \overline{\xi}_{\alpha, h}\gamma_{h}(u).
\end{equation*}
Typically, the constants $\underline{\xi}_{\alpha, h}$ and $\overline{\xi}_{\alpha, h}$ are chosen equal, in turn, it leads to
\begin{equation*}
-\xi_{\alpha, h} \gamma_h(u)\leq \widehat{\epsilon}_m(u) < \xi_{\alpha, h}\gamma_h(u),
\end{equation*}
where $\xi_{\alpha, h}$ is a \textit{single} tuning parameter. By the law of large numbers, one may achieve
\begin{equation*}
\text{Pr}[-\xi_{\alpha, h}\gamma_h(u)\leq \X_{n+h}(u) - \widehat{\X}_{n+h|n}(u) \leq \xi_{\alpha, h}\gamma_h(u)] \approx \frac{1}{M}\sum^{M}_{m=1}\mathds{1}[-\xi_{\alpha, h}\gamma_h(u)\leq \widehat{\epsilon}_m(u) \leq \xi_{\alpha, h}\gamma_h(u)],
\end{equation*} 
when $M$ is reasonably large, where $n$ denotes a jump-off year.

We summarize the split conformal prediction using Algorithm~1.

\begin{algorithm}[H]
\caption{Calibrate scaled residual band for forecast errors}\label{alg:calibrate_band}

\begin{algorithmic}[1]
\Require Validation curves $\{\X_m(u)\}_{m=1}^M$, and their forecasts $\{\widehat{\X}_m(u)\}_{m=1}^M$, significance level $\alpha$
\Ensure Pointwise summary statistic $\gamma_h(u)$ and tuning constant(s) $\xi_{\alpha, h}$ (or $(\underline{\xi}_{\alpha, h},\overline{\xi}_{\alpha, h})$)

\State \textbf{Compute residual curves:} $\widehat{\epsilon}_m(u) \gets \X_m(u) - \widehat{\X}_m(u)$ \Comment{for all $u$}
\State \textbf{Compute pointwise scale function $\gamma_h(u)$:}  Let $\bm{v}(u) \gets \{\widehat{\epsilon}_m(u)\}_{m=1}^M$, and 
\begin{equation*}
\gamma_h(u) \gets \mathrm{Stat}\big(\bm{v}(u)\big)
\end{equation*}
 \Comment{$\mathrm{Stat}\in\{\mathrm{sd},\mathrm{IQR},\mathrm{MAD}\}$}
    \Comment{If using quantile: $\gamma(u) \gets \widehat{Q}_{1-\alpha}\big(\{|\widehat{\epsilon}_m(u)|\}_{m=1}^M\big)$ and $\widehat{Q}$ denotes empirical quantile}

\State \textbf{Calibrate tuning parameter (symmetric case): $\xi_{\alpha, h}$}
\State \textbf{Define ECP}
\[
\text{ECP}(\xi_{\alpha,h}) \;=\; \frac{1}{M}\sum_{m=1}^M 
\mathds{1}\!\left[\,\forall u\in\mathcal{U}:\; -\xi_{\alpha,h}\,\gamma_h(u)\le \widehat{\epsilon}_m(u)\le \xi_{\alpha,h}\,\gamma_h(u)\right].
\]
\State $\xi_{\alpha,h} \gets \arg\min_{\xi_{\alpha,h} \ge 0}\{\xi_{\alpha,h}:\; \text{ECP}(\xi_{\alpha,h})\ge 1-\alpha\}$ \Comment{use a grid search over $\xi_{\alpha,h}$}

\State \Return $\gamma_h(u)$ and $\xi_{\alpha, h}$
\end{algorithmic}

\end{algorithm}

\subsection{Sequential conformal prediction}\label{sec:3.2}

Without the need for a validation set, this sequential conformal prediction can automatically tune the predictive quantiles of the absolute residuals as new data arrive. With the last 20 years as the test set, we use the other years to compute the absolute residuals $\{|\epsilon_1(u_j)|,\dots,|\epsilon_{\eta}(u_j)|\}$ for a given age~$u_j$. At the quantile of $1-\alpha$, we fit a quantile regression on lagged residuals, where the order of the autoregression AR($p$) is determined by an information criterion, such as the Akaike information criterion. Conditional on the most recent $p$ number of the absolute residuals as the input, we predict a one-step-ahead quantile, denoted as $\widehat{q}_{\eta+1, \alpha}(u_j)$. The prediction intervals are then given by
\begin{equation*}
\widehat{\X}_{\eta+1}(u_j) \pm \widehat{q}_{\eta+1, \alpha}(u_j).
\end{equation*}
Once the $\X_{\eta+1}(u_j)$ arrives, we can update the residual $|\epsilon_{\eta+1}(u_j)|$ and refit.

We summarize the sequential conformal prediction using Algorithm~2.

\begin{algorithm}[H]
\caption{Sequential conformal prediction via AR($p$) quantile regression (per age~$u_j$)}\label{alg:seq_conformal_ar_quantile}
\begin{algorithmic}[1]
\Require Age grid $\{u_j\}_{j=1}^J$, observed curves $\{\X_t(u_j)\}$, point forecasts $\{\widehat{\X}_t(u_j)\}$, level $\alpha$, test length $n_{\text{test}}=20$
\Ensure Sequential prediction intervals $\widehat{\X}_{\ell}(u_j)\pm \widehat{q}_{\ell,\alpha}(u_j)$ for $\ell$ in the test period

\State Set training end index $\eta \gets n - n_{\text{test}}$ \Comment{$\eta$ = last index before the final 20 years}

For $\ell=\eta+1,\dots,n$ and $j=1,\dots,J$, compute past absolute residuals:
        \[
        r_s(u_j) \gets \left|\epsilon_s(u_j)\right| = \left|\X_s(u_j)-\widehat{\X}_s(u_j)\right|,\quad s=1,\dots,\ell-1
        \]
        \State Select AR order $p \gets \arg\min_p \mathrm{IC}(p)$ (e.g., AIC) using $\{r_s(u_j)\}_{s=1}^{\ell-1}$
        \State Fit quantile regression at level $(1-\alpha)$:
        \[
        Q_{1-\alpha}\!\left[r_s(u_j)\mid r_{s-1}(u_j),\dots,r_{s-p}(u_j)\right],\quad s=p+1,\dots,\ell-1
        \]
        \State Predict one-step-ahead quantile using latest lags:
        \[
        \widehat{q}_{\ell,\alpha}(u_j) \gets \widehat{Q}_{1-\alpha}\!\left[r_{\ell}(u_j)\mid r_{\ell-1}(u_j),\dots,r_{\ell-p}(u_j)\right]
        \]
        \State Output prediction interval:
        \[
        \widehat{\X}_{\ell}(u_j)\ \pm\ \widehat{q}_{\ell,\alpha}(u_j)
        \]
    \State When $\X_{\ell}(u_j)$ is observed for all $j$, update residuals $r_{\ell}(u_j)=|\X_{\ell}(u_j)-\widehat{\X}_{\ell}(u_j)|$ and continue

\end{algorithmic}
\end{algorithm}

\vspace{.2in}

The sequential conformal prediction conformalizes the nonconformity scores sequentially, and uses an AR($p$) quantile regression to adapt the $(1-\alpha)$ quantile as data arrive, instead of relying on a calibration set. Under temporal dependence, we expect approximate marginal coverage near $(1-\alpha)$ under weak dependence and a reasonably specified quantile regression model, with possibly over or under-coverage under strong dependence or structural breaks.

\section{Evaluation of interval forecast accuracy}\label{sec:4}

\subsection{Expanding and rolling forecast schemes}\label{sec:4.1}

We divide the Australian age-specific mortality data into a training sample from 1921 to 1980, a validation sample from 1981 to 2001, and a test sample from 2002 to 2021. With a test sample of 20 years, we can evaluate forecast horizons from $h=1$ to 20, representing different degrees of forecast uncertainty. In the validation sample, there are 21 years, since some summary statistics, such as standard deviation, IQR, and MAD, require at least two years of data to compute.

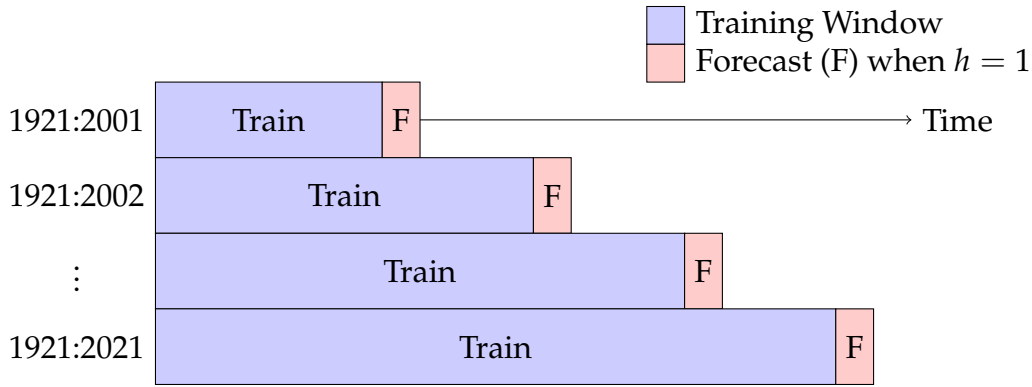
\begin{figure}[!htb]
\begin{center}
\begin{tikzpicture}
\draw[->] (0,0) -- (10,0) node[right] {Time};
    
\draw[fill=blue!20] (0,-0.5) rectangle (3,0.5) node[midway] {Train};
\draw[fill=red!20] (3,-0.5) rectangle (3.5,0.5) node[midway] {F};
    
\draw[fill=blue!20] (0,-1.5) rectangle (5,-0.5) node[midway] {Train};
\draw[fill=red!20] (5,-1.5) rectangle (5.5,-0.5) node[midway] {F};
    
\draw[fill=blue!20] (0,-2.5) rectangle (7,-1.5) node[midway] {Train};
\draw[fill=red!20] (7,-2.5) rectangle (7.5,-1.5) node[midway] {F};
    
\draw[fill=blue!20] (0,-3.5) rectangle (9,-2.5) node[midway] {Train};
\draw[fill=red!20] (9,-3.5) rectangle (9.5,-2.5) node[midway] {F};
    
\node[left] at (0,0) {1921:2001};
\node[left] at (0,-1) {1921:2002};
\node[left] at (0,-2) {\hspace{-0.8in}{$\vdots$}};
\node[left] at (0,-3) {1921:2021};
    
\draw[fill=blue!20] (6.5,1) rectangle (7,1.5);
\node[right] at (7,1.25) {Training Window};
\draw[fill=red!20] (6.5,0.5) rectangle (7,1);
\node[right] at (7,0.75) {Forecast (F) when $h=1$};
\end{tikzpicture}
\end{center}
\caption{A diagram of the expanding-window forecast scheme.}\label{fig:2}
\end{figure}

We examine the expanding and rolling forecast schemes. An expanding window analysis of a time-series model is commonly used to assess model and parameter stability over time. For instance, using the first 81 years from 1921 to 2001, we can produce one- to 20-step-ahead forecasts. Using an expanding window scheme, we estimate the parameters of the time-series forecasting models based on the first 82 years from 1921 to 2002. Forecasts from the estimated model are produced for one- to 19-step-ahead forecasts. We iterate this process by increasing the sample size by one year until we reach the end of the data period in 2021. This iterative process produces 20 one-step-ahead forecasts, 19 two-step-ahead forecasts, $\dots$, and one 20-step-ahead forecast. In Figure~\ref{fig:2}, we show a diagram of the expanding window forecast scheme for the forecast horizon $h=1$, although we also consider other forecast horizons from $h=2$ to 20.

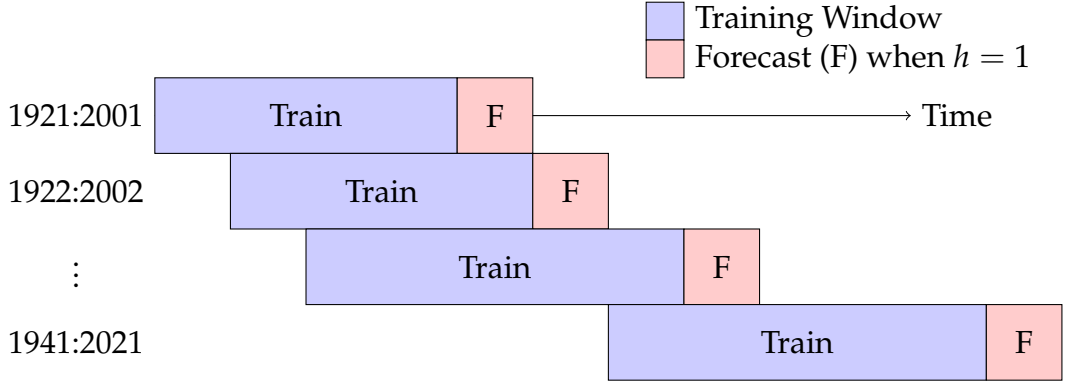
\begin{figure}[!htb]
\begin{center}
\begin{tikzpicture}
\draw[->] (0,0) -- (10,0) node[right] {Time};
    
\draw[fill=blue!20] (0,-0.5) rectangle (4,0.5) node[midway] {Train};
\draw[fill=red!20] (4,-0.5) rectangle (5,0.5) node[midway] {F};
    
\draw[fill=blue!20] (1,-1.5) rectangle (5,-0.5) node[midway] {Train};
\draw[fill=red!20] (5,-1.5) rectangle (6,-0.5) node[midway] {F};
    
\draw[fill=blue!20] (2,-2.5) rectangle (7,-1.5) node[midway] {Train};
\draw[fill=red!20] (7,-2.5) rectangle (8,-1.5) node[midway] {F};
    
\draw[fill=blue!20] (6,-3.5) rectangle (11,-2.5) node[midway] {Train};
\draw[fill=red!20] (11,-3.5) rectangle (12,-2.5) node[midway] {F};
    
\node[left] at (0,0) {1921:2001};
\node[left] at (0,-1) {1922:2002};
\node[left] at (0,-2) {\hspace{-0.8in}{$\vdots$}};
\node[left] at (0,-3) {1941:2021};
    
\draw[fill=blue!20] (6.5,1) rectangle (7,1.5);
\node[right] at (7,1.25) {Training Window};
\draw[fill=red!20] (6.5,0.5) rectangle (7,1);
\node[right] at (7,0.75) {Forecast (F) when $h=1$};
\end{tikzpicture}
\end{center}
\caption{A diagram of the rolling-window forecast scheme.}\label{fig:3}
\end{figure}

A rolling-window analysis of a time-series model is an alternative forecasting scheme that sequentially removes data from the distant past, thereby allowing gradual changes. For instance, using the first 81 years from 1921 to 2001, we can produce one- to 20-step-ahead forecasts. Using a rolling window scheme, we estimate the parameters of the time-series forecasting models for the period 1922 to 2002. Forecasts from the estimated model are produced for one- to 19-step-ahead forecasts. We iterate on this process by adjusting the sample size until we reach the end of the data period in 2021. In Figure~\ref{fig:3}, we show a diagram of the expanding window forecast scheme for the forecast horizon $h=1$.

\subsection{Measure of interval forecast accuracy}\label{sec:4.2}

To evaluate interval forecast accuracy, we consider the ECP, the coverage probability difference (CPD), and the interval score of \cite{GR07}. For each year in the forecast period, the $h$-step-ahead prediction intervals are calculated at the $(1-\alpha)$ nominal coverage probability. The lower and upper bounds, denoted by $\left[\widehat{\X}_{m+\vartheta}^{\text{lb}}(u), \widehat{\X}_{m+\vartheta}^{\text{ub}}(u)\right]$, are not required to be centered around the point forecasts. The ECP and CPD are defined as
\begin{align*}
\text{ECP}_h &= \frac{1}{101\times (21-h)}\sum^{20}_{\vartheta=h}\sum^{101}_{u=1}\mathds{1}\left\{\widehat{\X}_{m+\vartheta}^{\text{lb}}(u) \leq \X_{m+\vartheta}(u)\leq \widehat{\X}_{m+\vartheta}^{\text{ub}}(u)\right\}, \\
\text{CPD}_h &= \left|\frac{1}{101\times (21-h)}\sum^{20}_{\vartheta=h}\sum^{101}_{u=1}\left[\mathds{1}\left\{\X_{m+\vartheta}(u) > \widehat{\X}_{m+\vartheta}^{\text{ub}}(u) \right\} + \mathds{1}\left\{\X_{m+\vartheta}(u) < \widehat{\X}_{m+\vartheta}^{\text{lb}}(u)\right\}\right]-\alpha\right|,
\end{align*}
where $\alpha$ denotes a level of significance, customarily $\alpha=0.2$ or 0.05, and $\mathds{1}\{\cdot\}$ represents binary indicator function.

The ECP assesses the coverage, but it cannot evaluate the sharpness of a prediction interval. By combining the coverage and sharpness, we consider a scoring rule for the prediction interval at age $u$, defined as
\begin{align*}
S_{\alpha,\vartheta}[\widehat{\X}_{m+\vartheta}^{\text{lb}}(u_j), \widehat{\X}_{m+\vartheta}^{\text{ub}}(u_j), \X_{m+\vartheta}(u_j)] =& \left[\widehat{\X}_{m+\vartheta}^{\text{ub}}(u_j) - \widehat{\X}_{m+\vartheta}^{\text{lb}}(u_j)\right] \\
&+\frac{2}{\alpha}[\widehat{\X}_{m+\vartheta}^{\text{lb}}(u_j) - \X_{m+\vartheta}(u_j)]\mathds{1}\left\{\X_{m+\vartheta}(u_j) < \widehat{\X}_{m+\vartheta}^{\text{lb}}(u_j)\right\} \\
&+\frac{2}{\alpha}[\X_{m+\vartheta}(u_j) - \widehat{\X}_{m+\vartheta}^{\text{ub}}(u_j)]\mathds{1}\left\{\X_{m+\vartheta}(u_j) > \widehat{\X}_{m+\vartheta}^{\text{ub}}(u_j)\right\}.
\end{align*}
The mean interval score is given as 
\begin{equation*}
\overline{S}_{\alpha,h} = \frac{1}{101\times(21-h)}\sum^{20}_{\vartheta=h}\sum^{101}_{j=1}S_{\alpha,\vartheta}[\widehat{\X}_{m+\vartheta}^{\text{lb}}(u_j), \widehat{\X}_{m+\vartheta}^{\text{ub}}(u_j), \X_{m+\vartheta}(u_j)].
\end{equation*}
Given the same ECP, the mean interval score rewards a narrow prediction interval width.

\subsection{Comparison between the expanding and rolling windows}\label{sec:4.3}

At the nominal coverage probabilities of 80\% and 95\%, we evaluate and compare the interval forecast accuracy in Table~\ref{tab:2}, as measured by the ECP, CPD, and mean interval score $\overline{S}_{\alpha}$, using the expanding- and rolling-window forecast schemes. For both female and male data, the difference between the two schemes is marginal. 
\begin{center}
\tabcolsep 0.075in
\renewcommand{\arraystretch}{0.72}
\begin{small}
\begin{longtable}{@{}llllrrrrrrrr}
\caption{\small At the nominal coverage probabilities of 80\% and 95\%, we compare the interval forecast accuracy, as measured by the ECP, CPD, and mean interval score, over the 20 forecast horizons. Based on the summary statistics, we compare the differences between the expanding- and rolling-window forecast schemes using the functional time-series forecasting method with $K=6$.}\label{tab:2}\\
  \toprule
&  	&	&  & \multicolumn{4}{c}{Summary statistics (Female)} & \multicolumn{4}{c}{Summary statistics (Male)} \\
	\cmidrule(lr){5-8}\cmidrule(lr){9-12}
$\alpha$ & Scheme & Metric & Summary & Quant & sd  & IQR & MAD  & Quant & sd  & IQR & MAD \\
  \midrule
0.2 	& Rolling 	&  ECP 	& Min. 	& 0.693 & 0.802 & 0.782 & 0.782 & 0.738 & 0.798 & 0.792 & 0.762 \\ 
	& 		& 		& 1st.Qu 	& 0.802 & 0.818 & 0.805 & 0.814 & 0.763 & 0.807 & 0.807 & 0.804 \\ 
 	& 		& 		& \cellcolor{yellow}Median & \cellcolor{yellow} 0.808 & \cellcolor{yellow} 0.828 & \cellcolor{yellow} 0.827 & \cellcolor{yellow} 0.835 & \cellcolor{yellow} 0.778 & \cellcolor{yellow} 0.868 & \cellcolor{yellow} 0.856 & \cellcolor{yellow} 0.847 \\ 
 	& 		& 		&  \cellcolor{skyblue}Mean & \cellcolor{skyblue}0.807 & \cellcolor{skyblue}0.829 & \cellcolor{skyblue}0.829 & \cellcolor{skyblue}0.832 & \cellcolor{skyblue}0.776 & \cellcolor{skyblue}0.866 & \cellcolor{skyblue}0.853 & \cellcolor{skyblue}0.845 \\ 
 	& 		& 		& 3rd.Qu & 0.822 & 0.840 & 0.850 & 0.844 & 0.792 & 0.912 & 0.892 & 0.882 \\ 
 	& 		& 		& Max & 0.845 & 0.856 & 0.882 & 0.886 & 0.806 & 0.949 & 0.934 & 0.919 \\ 
  \cmidrule{3-12}
 	&  		& CPD 	&  Min. & 0.001 & 0.002 & 0.001 & 0.002 & 0.000 & 0.000 & 0.002 & 0.001 \\ 
	&  		& 		& 1st.Qu. & 0.004 & 0.018 & 0.017 & 0.017 & 0.008 & 0.007 & 0.008 & 0.014 \\ 
	&  		& 		& \cellcolor{yellow}Median & \cellcolor{yellow} 0.011 & \cellcolor{yellow} 0.028 & \cellcolor{yellow} 0.027 & \cellcolor{yellow} 0.035 & \cellcolor{yellow} 0.022 & \cellcolor{yellow} 0.068 & \cellcolor{yellow} 0.056 & \cellcolor{yellow} 0.050 \\ 
	&  		& 		& \cellcolor{skyblue}Mean & \cellcolor{skyblue}0.019 & \cellcolor{skyblue}0.029 & \cellcolor{skyblue}0.032 & \cellcolor{skyblue}0.034 & \cellcolor{skyblue}0.025 & \cellcolor{skyblue}0.066 & \cellcolor{skyblue}0.056 & \cellcolor{skyblue}0.051 \\ 
	&  		& 		& 3rd.Qu. & 0.023 & 0.040 & 0.050 & 0.044 & 0.037 & 0.112 & 0.092 & 0.082 \\ 
	&  		& 		& Max. & 0.107 & 0.056 & 0.082 & 0.086 & 0.062 & 0.149 & 0.134 & 0.119 \\ 
\cmidrule{3-12}
	& 		& $\overline{S}_{0.2}$ & Min. & 0.491 & 0.488 & 0.488 & 0.490 & 0.413 & 0.416 & 0.437 & 0.431 \\ 
	 & 		& 		& 1st.Qu & 0.579 & 0.572 & 0.603 & 0.603 & 0.566 & 0.585 & 0.614 & 0.614 \\ 
 & 	& & \cellcolor{yellow}Median & \cellcolor{yellow}0.700 & \cellcolor{yellow}0.693 & \cellcolor{yellow}0.723 & \cellcolor{yellow}0.733 & \cellcolor{yellow}0.765 & \cellcolor{yellow}0.820 & \cellcolor{yellow}0.888 & \cellcolor{yellow}0.909 \\ 
 & 	& & \cellcolor{skyblue}Mean & \cellcolor{skyblue}0.747 & \cellcolor{skyblue}0.842 & \cellcolor{skyblue}0.902 & \cellcolor{skyblue}0.964 & \cellcolor{skyblue}0.773 & \cellcolor{skyblue}1.037 & \cellcolor{skyblue}1.112 & \cellcolor{skyblue}1.181 \\ 
 & 	& & 3rd.Qu. & 0.897 & 0.956 & 1.026 & 1.096 & 0.993 & 1.023 & 1.176 & 1.250 \\ 
 & 	& & Max & 1.234 & 2.492 & 2.479 & 2.490 & 1.146 & 3.988 & 3.983 & 5.145 \\   	
\cmidrule{2-12} 
 & Expanding & ECP & Min & 0.663 & 0.748 & 0.748 & 0.779 & 0.743 & 0.797 & 0.809 & 0.803 \\ 
 & & & 1st.Qu & 0.781 & 0.816 & 0.807 & 0.813 & 0.784 & 0.854 & 0.851 & 0.847 \\ 
 & & & \cellcolor{yellow}Median & \cellcolor{yellow}0.789 & \cellcolor{yellow}0.822 & \cellcolor{yellow}0.825 & \cellcolor{yellow}0.820 & \cellcolor{yellow}0.797 & \cellcolor{yellow}0.885 & \cellcolor{yellow}0.871 & \cellcolor{yellow}0.871 \\ 
 & & &  \cellcolor{skyblue}Mean & \cellcolor{skyblue}0.783 & \cellcolor{skyblue}0.818 & \cellcolor{skyblue}0.819 & \cellcolor{skyblue}0.823 & \cellcolor{skyblue}0.796 & \cellcolor{skyblue}0.873 & \cellcolor{skyblue}0.867 & \cellcolor{skyblue}0.868 \\ 
 & & &  3rd.Qu & 0.796 & 0.826 & 0.833 & 0.837 & 0.812 & 0.897 & 0.881 & 0.891 \\ 
 & & &  Max & 0.809 & 0.861 & 0.861 & 0.861 & 0.836 & 0.917 & 0.931 & 0.919 \\ 
\cmidrule{3-12}
& & CPD & Min & 0.002 & 0.002 & 0.001 & 0.001 & 0.001 & 0.002 & 0.009 & 0.003 \\ 
& & & 1st.Qu. & 0.006 & 0.019 & 0.015 & 0.014 & 0.007 & 0.054 & 0.051 & 0.047 \\ 
& & &  \cellcolor{yellow}Median & \cellcolor{yellow}0.011 & \cellcolor{yellow}0.023 & \cellcolor{yellow}0.027 & \cellcolor{yellow}0.022 & \cellcolor{yellow}0.013 & \cellcolor{yellow}0.085 & \cellcolor{yellow}0.071 & \cellcolor{yellow}0.071 \\ 
& & &  \cellcolor{skyblue}Mean & \cellcolor{skyblue}0.019 & \cellcolor{skyblue}0.026 & \cellcolor{skyblue}0.027 & \cellcolor{skyblue}0.026 & \cellcolor{skyblue}0.019 & \cellcolor{skyblue}0.073 & \cellcolor{skyblue}0.067 & \cellcolor{skyblue}0.068 \\ 
& & &  3rd.Qu. & 0.019 & 0.031 & 0.034 & 0.037 & 0.028 & 0.097 & 0.081 & 0.091 \\ 
& & &  Max & 0.137 & 0.061 & 0.061 & 0.061 & 0.057 & 0.117 & 0.131 & 0.119 \\   
\cmidrule{3-12}
& & $\overline{S}_{0.2}$ & Min & 0.501 & 0.494 & 0.489 & 0.495 & 0.407 & 0.413 & 0.428 & 0.420 \\ 
& & & 1st.Qu & 0.577 & 0.570 & 0.592 & 0.594 & 0.567 & 0.577 & 0.624 & 0.630 \\ 
& & &  \cellcolor{yellow}Median & \cellcolor{yellow}0.722 & \cellcolor{yellow}0.700 & \cellcolor{yellow}0.769 & \cellcolor{yellow}0.757 & \cellcolor{yellow}0.781 & \cellcolor{yellow}0.802 & \cellcolor{yellow}0.820 & \cellcolor{yellow}0.871 \\ 
& & &  \cellcolor{skyblue}Mean & \cellcolor{skyblue}0.749 & \cellcolor{skyblue}0.831 & \cellcolor{skyblue}0.881 & \cellcolor{skyblue}0.935 & \cellcolor{skyblue}0.772 & \cellcolor{skyblue}1.014 & \cellcolor{skyblue}1.117 & \cellcolor{skyblue}1.342 \\ 
& & &  3rd.Qu & 0.884 & 0.922 & 0.992 & 1.013 & 0.981 & 0.971 & 1.029 & 1.061 \\ 
& & &  Max & 1.227 & 2.406 & 2.359 & 2.370 & 1.126 & 4.665 & 4.716 & 6.381 \\   
\midrule  
0.05 & Rolling & ECP & Min & 0.832 & 0.928 & 0.914 & 0.908 & 0.767 & 0.929 & 0.925 & 0.762 \\ 
&  & & 1st.Qu & 0.888 & 0.944 & 0.939 & 0.942 & 0.876 & 0.939 & 0.937 & 0.936 \\ 
&  & & \cellcolor{yellow}Median & \cellcolor{yellow}0.911 & \cellcolor{yellow}0.951 & \cellcolor{yellow}0.948 & \cellcolor{yellow}0.947 & \cellcolor{yellow}0.893 & \cellcolor{yellow}0.963 & \cellcolor{yellow}0.950 & \cellcolor{yellow}0.956 \\ 
&  & & \cellcolor{skyblue}Mean & \cellcolor{skyblue}0.906 & \cellcolor{skyblue}0.952 & \cellcolor{skyblue}0.949 & \cellcolor{skyblue}0.948 & \cellcolor{skyblue}0.884 & \cellcolor{skyblue}0.960 & \cellcolor{skyblue}0.956 & \cellcolor{skyblue}0.949 \\ 
&  & & 3rd.Qu & 0.925 & 0.957 & 0.957 & 0.959 & 0.902 & 0.979 & 0.973 & 0.975 \\ 
&  & & Max & 0.980 & 0.990 & 0.985 & 0.976 & 1.000 & 0.993 & 0.996 & 0.996 \\ 
\cmidrule{3-12}
&  & CPD &  Min & 0.019 & 0.001 & 0.001 & 0.000 & 0.037 & 0.001 & 0.000 & 0.004 \\ 
&  & & 1st.Qu & 0.025 & 0.003 & 0.005 & 0.006 & 0.050 & 0.012 & 0.010 & 0.013 \\ 
&  & & \cellcolor{yellow}Median & \cellcolor{yellow}0.039 & \cellcolor{yellow}0.007 & \cellcolor{yellow}0.010 & \cellcolor{yellow}0.010 & \cellcolor{yellow}0.057 & \cellcolor{yellow}0.020 & \cellcolor{yellow}0.017 & \cellcolor{yellow}0.018 \\ 
&  & & \cellcolor{skyblue}Mean & \cellcolor{skyblue}0.047 & \cellcolor{skyblue}0.011 & \cellcolor{skyblue}0.014 & \cellcolor{skyblue}0.013 & \cellcolor{skyblue}0.071 & \cellcolor{skyblue}0.021 & \cellcolor{skyblue}0.019 & \cellcolor{skyblue}0.029 \\ 
&  & & 3rd.Qu & 0.062 & 0.012 & 0.019 & 0.016 & 0.074 & 0.029 & 0.027 & 0.027 \\ 
&  & & Max & 0.118 & 0.040 & 0.036 & 0.042 & 0.183 & 0.043 & 0.046 & 0.188 \\   
\cmidrule{3-12}
& 	& $\overline{S}_{0.05}$ & Min & 0.848 & 0.783 & 0.759 & 0.763 & 0.652 & 0.595 & 0.668 & 0.648 \\ 
&  	& & 1st.Qu & 0.923 & 0.860 & 0.926 & 0.952 & 0.820 & 0.801 & 0.891 & 0.903 \\ 
&  	& & \cellcolor{yellow}Median & \cellcolor{yellow}1.124 & \cellcolor{yellow}1.072 & \cellcolor{yellow}1.170 & \cellcolor{yellow}1.220 & \cellcolor{yellow}1.137 & \cellcolor{yellow}1.091 & \cellcolor{yellow}1.235 & \cellcolor{yellow}1.383 \\ 
&  	& & \cellcolor{skyblue}Mean & \cellcolor{skyblue}1.464 & \cellcolor{skyblue}1.710 & \cellcolor{skyblue}1.889 & \cellcolor{skyblue}2.030 & \cellcolor{skyblue}2.122 & \cellcolor{skyblue}2.093 & \cellcolor{skyblue}2.299 & \cellcolor{skyblue}3.639 \\ 
&  	& & 3rd.Qu & 1.419 & 1.486 & 1.796 & 2.087 & 1.417 & 1.398 & 1.763 & 2.140 \\ 
&  	& & Max & 6.793 & 9.916 & 9.956 & 8.972 & 21.150 & 16.992 & 17.952 & 42.663 \\ 
\cmidrule{2-12}
 & Expanding & ECP &  Min & 0.827 & 0.924 & 0.932 & 0.926 & 0.844 & 0.797 & 0.937 & 0.938 \\ 
 & & & 1st.Qu & 0.871 & 0.941 & 0.941 & 0.941 & 0.886 & 0.959 & 0.957 & 0.961 \\ 
 & & & \cellcolor{yellow}Median & \cellcolor{yellow}0.900 & \cellcolor{yellow}0.946 & \cellcolor{yellow}0.946 & \cellcolor{yellow}0.946 & \cellcolor{yellow}0.908 & \cellcolor{yellow}0.976 & \cellcolor{yellow}0.970 & \cellcolor{yellow}0.970 \\ 
 & & & \cellcolor{skyblue}Mean & \cellcolor{skyblue}0.892 & \cellcolor{skyblue}0.946 & \cellcolor{skyblue}0.948 & \cellcolor{skyblue}0.949 & \cellcolor{skyblue}0.906 & \cellcolor{skyblue}0.962 & \cellcolor{skyblue}0.967 & \cellcolor{skyblue}0.966 \\ 
 & & & 3rd.Qu & 0.911 & 0.949 & 0.953 & 0.957 & 0.916 & 0.981 & 0.975 & 0.972 \\ 
 & & & Max & 0.960 & 0.975 & 0.975 & 0.972 & 1.000 & 0.986 & 1.000 & 0.983 \\ 
\cmidrule{3-12}
& & CPD & Min & 0.010 & 0.000 & 0.001 & 0.002 & 0.019 & 0.000 & 0.004 & 0.007 \\ 
& & & 1st.Qu & 0.039 & 0.003 & 0.004 & 0.005 & 0.039 & 0.015 & 0.009 & 0.012 \\ 
& & & \cellcolor{yellow}Median & \cellcolor{yellow}0.050 & \cellcolor{yellow}0.007 & \cellcolor{yellow}0.009 & \cellcolor{yellow}0.008 & \cellcolor{yellow}0.043 & \cellcolor{yellow}0.027 & \cellcolor{yellow}0.020 & \cellcolor{yellow}0.020 \\ 
& & & \cellcolor{skyblue}Mean & \cellcolor{skyblue}0.059 & \cellcolor{skyblue}0.008 & \cellcolor{skyblue}0.009 & \cellcolor{skyblue}0.010 & \cellcolor{skyblue}0.054 & \cellcolor{skyblue}0.030 & \cellcolor{skyblue}0.019 & \cellcolor{skyblue}0.019 \\ 
& & & 3rd.Qu & 0.079 & 0.010 & 0.012 & 0.014 & 0.064 & 0.033 & 0.025 & 0.022 \\ 
& & & Max & 0.123 & 0.026 & 0.025 & 0.024 & 0.106 & 0.153 & 0.050 & 0.033 \\  
\cmidrule{3-12}
& & $\overline{S}_{0.05}$ & Min & 0.864 & 0.791 & 0.760 & 0.770 & 0.625 & 0.589 & 0.639 & 0.616 \\ 
& & & 1st.Qu & 0.921 & 0.856 & 0.915 & 0.939 & 0.791 & 0.798 & 0.912 & 0.934 \\ 
& & & \cellcolor{yellow}Median & \cellcolor{yellow}1.159 & \cellcolor{yellow}1.094 & \cellcolor{yellow}1.272 & \cellcolor{yellow}1.278 & \cellcolor{yellow}1.123 & \cellcolor{yellow}1.142 & \cellcolor{yellow}1.261 & \cellcolor{yellow}1.328 \\ 
& & & \cellcolor{skyblue}Mean & \cellcolor{skyblue}1.333 & \cellcolor{skyblue}1.591 & \cellcolor{skyblue}1.725 & \cellcolor{skyblue}1.981 & \cellcolor{skyblue}2.527 & \cellcolor{skyblue}2.172 & \cellcolor{skyblue}2.467 & \cellcolor{skyblue}3.532 \\ 
& & & 3rd.Qu & 1.428 & 1.405 & 1.700 & 1.902 & 1.391 & 1.319 & 1.640 & 1.785 \\ 
& & & Max & 3.834 & 8.249 & 7.779 & 8.613 & 20.847 & 21.365 & 19.233 & 24.418 \\ 
  \bottomrule
\end{longtable}
\end{small}
\end{center}

\vspace{-.4in}

In Table~\ref{tab:tune_1}, we present the selected tuning parameters $\xi_{\alpha}$ for different forecast horizons. Instead of $h=1$ to 20, we report $h=1,5,10,15,20$. As $h$ increases, the selected tuning parameters increase to reflect greater uncertainty in the point forecast. For a large horizon $h=20$, because of a relatively smaller number of observations for calibration, the selected tuning parameters can be unstably large. Possible mitigation strategies include 
\begin{inparaenum}
\item[(i)] smooth $\xi_{\alpha, h}$ across horizons using a low-order spline, LOESS \citep{Cleveland79} or a simple moving average, borrowing strength from neighboring horizons; 
\item[(ii)] impose mild shape constraints such as nondecreasing $\xi_{\alpha, h}$ in $h$ via isotonic regression, reflecting that uncertainty typically increases with horizon; and 
\item[(iii)] apply shrinkage by letting $\xi_{\alpha, h}$ towards the mean of $\xi_{\alpha, h}$. 
\end{inparaenum}
These steps reduce variability of the calibrated CPD for large~$h$. When the summary statistics is quantile, it can adaptive to the increase in forecast uncertainty resulting in the selected tuning parameter being around one.
\begin{center}
\tabcolsep 0.11in
\renewcommand{\arraystretch}{0.85}
\begin{small}
\begin{longtable}{@{}lllrrrrrrrr}
\caption{\small At the nominal coverage probabilities of 80\% and 95\%, we present the selected tuning parameter $\xi_{\alpha,h}$ at $h=1,5,10,15$ and $20$, due to the limited space. Based on the summary statistics, we compare the difference between the expanding- and rolling-window forecast schemes using the functional time-series forecasting method with $K=6$.}\label{tab:tune_1}\\
  \toprule
	&  	&	  & \multicolumn{4}{c}{Summary statistics (Female)} & \multicolumn{4}{c}{Summary statistics (Male)} \\
	\cmidrule(lr){4-7}\cmidrule(lr){8-11}
$\alpha$ & Scheme & $h$ &  Quant & sd  & IQR & MAD  & Quant & sd  & IQR & MAD \\
\midrule
0.2 	& rolling 	& 1 	& 1.000 & 1.350 	& 1.062 	& 1.400 	& 1.000 & 1.471 	& 1.206 	& 1.535 \\ 
  	& 		& 5 	& 1.020 & 1.564 	& 1.277 	& 1.678 	& 1.016 & 1.825 	& 1.484 	& 1.969 \\ 
    	& 		& 10 & 1.020 & 1.819 	& 1.551 	& 2.000 	& 1.016 & 2.606 	& 2.356 	& 3.188 \\ 
    	& 		& 15 & 1.045 & 2.103 	& 1.788 	& 2.325 	& 1.031 & 3.050 	& 2.913 	& 4.850 \\ 
    	& 		& 20 & 1.095 & 10.600 	& 14.900 	& 10.100 	& 1.069 & 23.300 	& 32.900 	& 9.718 \\ 
\cmidrule{2-11}
	& expanding 	& 1 & 1.000 & 1.344 & 1.083 & 1.441 & 1.000 & 1.475 & 1.194 & 1.539 \\ 
	& 			& 5 & 1.023 & 1.536 & 1.286 & 1.688 & 1.016 & 1.932 & 1.825 & 2.527 \\ 
	& 			& 10 & 1.021 & 1.894 & 1.738 & 2.244 & 1.015 & 2.450 & 1.694 & 2.909 \\ 
  	& 			& 15 & 1.045 & 2.131 & 1.925 & 2.575 & 1.023 & 2.456 & 1.678 & 2.444 \\ 
  	& 			& 20 & 1.091 & 10.500 & 14.500 & 9.700 & 1.056 & 29.700 & 42.500 & 29.700 \\       
\midrule
0.05 & rolling  	& 1 	& 1.000 & 2.063 & 1.775 & 2.325 & 1.000 & 2.188 & 2.000 & 2.491 \\ 
   	& 		&  5 	& 1.031 & 2.338 & 2.125 & 2.825 & 1.020 & 2.660 & 2.325 & 3.150 \\ 
	& 		&   10 & 1.056 & 2.688 & 2.388 & 3.275 & 1.049 & 3.444 & 3.350 & 4.725 \\ 
  	& 		&   15 & 1.063 & 3.588 & 3.384 & 5.350 & 1.054 & 4.225 & 4.350 & 10.300 \\ 
  	& 		&   20 & 10.901 & 45.700 & 64.900 & 39.300 & 25.752 & 103.300 & 154.500 & 9.718 \\ 
\cmidrule{2-11}
  	& expanding 	& 1 & 1.000 & 2.081 & 1.783 & 2.350 & 1.000 & 2.194 & 1.873 & 2.419 \\ 
  	& 			& 5 & 1.025 & 2.313 & 2.150 & 2.825 & 1.028 & 2.966 & 2.900 & 3.969 \\ 
  	& 			& 10 & 1.056 & 2.900 & 2.763 & 3.683 & 1.034 & 3.550 & 2.731 & 4.562 \\ 
  	& 			& 15 & 1.069 & 3.394 & 3.450 & 4.900 & 1.044 & 3.400 & 2.700 & 4.500 \\ 
  	& 			& 20 & 5.950 & 39.300 & 52.100 & 39.300 & 25.752 & 141.700 & 180.100 & 128.900 \\ 
\bottomrule
\end{longtable}
\end{small}
\end{center}

\vspace{-.5in}

\subsection{Comparison between the number of retained components}

Using the expanding-window forecast scheme (Table~\ref{tab:3}), we evaluate the interval forecast accuracy of the functional time-series forecasting method with two component selection methods: EVR and $K=6$. When the number of components is determined by the EVR criterion, the selected $K$ tends to be smaller than $K=6$, leading to larger residuals. Despite the potential for under-fitting, it only affects the quantile-based interval forecast accuracy, especially for the male population.
\begin{center}
\renewcommand{\arraystretch}{0.68}
\begin{small}
\tabcolsep 0.09in
\begin{longtable}{@{}llllrrrrrrrr}
\caption{\small At the nominal coverage probabilities of 80\% and 95\%, we compare the interval forecast accuracy, as measured by the ECP, CPD, and mean interval score, across the 20 forecast horizons. Based on the summary statistics, we present results from the functional time-series forecasting method, with $K$ determined via the EVR criterion.}\label{tab:3}\\
  \toprule
 & &  		&  & \multicolumn{4}{c}{Summary statistics (Female)} & \multicolumn{4}{c}{Summary statistics (Male)} \\
	\cmidrule(lr){5-8}\cmidrule(lr){9-12}
$K$ & $\alpha$ & Metric & Summary & Quant & sd  & IQR & MAD  & Quant & sd  & IQR & MAD \\
  \midrule
EVR & 0.2 & ECP & Min & 0.756 & 0.779 & 0.545 & 0.683 & 0.614 & 0.785 & 0.792 & 0.800 \\ 
 	& &	& 1st.Qu & 0.782 & 0.809 & 0.816 & 0.808 & 0.655 & 0.856 & 0.844 & 0.836 \\ 
 	& &	& \cellcolor{yellow}Median & \cellcolor{yellow}0.788 & \cellcolor{yellow}0.830 & \cellcolor{yellow}0.832 & \cellcolor{yellow}0.839 & \cellcolor{yellow}0.662 & \cellcolor{yellow}0.869 & \cellcolor{yellow}0.862 & \cellcolor{yellow}0.858 \\ 
 	& &	& \cellcolor{skyblue}Mean & \cellcolor{skyblue}0.791 & \cellcolor{skyblue}0.829 & \cellcolor{skyblue}0.821 & \cellcolor{skyblue}0.827 & \cellcolor{skyblue}0.661 & \cellcolor{skyblue}0.863 & \cellcolor{skyblue}0.854 & \cellcolor{skyblue}0.853 \\ 
 	& &	& 3rd.Qu & 0.802 & 0.853 & 0.852 & 0.855 & 0.674 & 0.886 & 0.872 & 0.872 \\ 
	& &	&  Max & 0.822 & 0.870 & 0.881 & 0.883 & 0.680 & 0.909 & 0.900 & 0.903 \\ 
\cmidrule{3-12}	
	&	& CPD & Min. & 0.002 & 0.002 & 0.003 & 0.003 & 0.120 & 0.001 & 0.008 & 0.000 \\ 
  	&	& & 1st.Qu & 0.005 & 0.014 & 0.024 & 0.020 & 0.126 & 0.056 & 0.044 & 0.036 \\ 
 	&	& & \cellcolor{yellow}Median & \cellcolor{yellow}0.014 & \cellcolor{yellow}0.030 & \cellcolor{yellow}0.035 & \cellcolor{yellow}0.042 & \cellcolor{yellow}0.138 & \cellcolor{yellow}0.069 & \cellcolor{yellow}0.062 & \cellcolor{yellow}0.058 \\ 
 	&	& & \cellcolor{skyblue}Mean & \cellcolor{skyblue}0.015 & \cellcolor{skyblue}0.033 & \cellcolor{skyblue}0.048 & \cellcolor{skyblue}0.043 & \cellcolor{skyblue}0.139 & \cellcolor{skyblue}0.065 & \cellcolor{skyblue}0.055 & \cellcolor{skyblue}0.053 \\ 
 	&	& & 3rd.Qu & 0.018 & 0.053 & 0.060 & 0.057 & 0.145 & 0.086 & 0.072 & 0.072 \\ 
	&	& & Max & 0.044 & 0.070 & 0.255 & 0.117 & 0.186 & 0.109 & 0.100 & 0.103 \\ 
\cmidrule{3-12}  
	&	& $\overline{S}_{0.2}$ & Min & 0.715 & 0.898 & 0.902 & 0.960 & 0.766 & 0.953 & 0.982 & 0.987 \\ 
 	&	& & 1st.Qu & 0.801 & 1.115 & 1.155 & 1.133 & 0.917 & 1.285 & 1.330 & 1.317 \\ 
	&	& &  \cellcolor{yellow}Median & \cellcolor{yellow}0.908 & \cellcolor{yellow}1.407 & \cellcolor{yellow}1.462 & \cellcolor{yellow}1.501 & \cellcolor{yellow}1.094 & \cellcolor{yellow}1.611 & \cellcolor{yellow}1.603 & \cellcolor{yellow}1.716 \\ 
 	&	& & \cellcolor{skyblue}Mean & \cellcolor{skyblue}0.925 & \cellcolor{skyblue}1.680 & \cellcolor{skyblue}1.628 & \cellcolor{skyblue}1.810 & \cellcolor{skyblue}1.082 & \cellcolor{skyblue}1.904 & \cellcolor{skyblue}1.984 & \cellcolor{skyblue}2.077 \\ 
	&	& & 3rd.Qu & 1.054 & 1.791 & 1.867 & 1.831 & 1.251 & 2.090 & 2.112 & 2.338 \\ 
	&	& & Max & 1.189 & 4.667 & 3.241 & 5.735 & 1.341 & 4.700 & 4.655 & 4.740 \\ 
\midrule
& 0.05 & ECP &  Min & 0.832 & 0.923 & 0.545 & 0.683 & 0.726 & 0.927 & 0.932 & 0.936 \\ 
	& &	&  1st.Qu & 0.874 & 0.931 & 0.941 & 0.939 & 0.761 & 0.960 & 0.957 & 0.954 \\ 
	& &	&  \cellcolor{yellow}Median & \cellcolor{yellow}0.893 & \cellcolor{yellow}0.947 & \cellcolor{yellow}0.955 & \cellcolor{yellow}0.948 & \cellcolor{yellow}0.797 & \cellcolor{yellow}0.971 & \cellcolor{yellow}0.964 & \cellcolor{yellow}0.962 \\ 
	& &	&  \cellcolor{skyblue}Mean & \cellcolor{skyblue}0.893 & \cellcolor{skyblue}0.947 & \cellcolor{skyblue}0.934 & \cellcolor{skyblue}0.937 & \cellcolor{skyblue}0.807 & \cellcolor{skyblue}0.967 & \cellcolor{skyblue}0.964 & \cellcolor{skyblue}0.962 \\ 
	& &	&  3rd.Qu & 0.902 & 0.955 & 0.965 & 0.963 & 0.820 & 0.978 & 0.972 & 0.973 \\ 
	& &	&  Max & 1.000 & 0.985 & 0.980 & 0.970 & 1.000 & 0.985 & 0.990 & 0.988 \\ 
\cmidrule{3-12}
	& & CPD & Min & 0.045 & 0.001 & 0.003 & 0.000 & 0.050 & 0.007 & 0.002 & 0.000 \\ 
	& &	&  1st.Qu & 0.050 & 0.004 & 0.008 & 0.006 & 0.130 & 0.013 & 0.009 & 0.008 \\ 
	& &	&  \cellcolor{yellow}Median & \cellcolor{yellow}0.057 & \cellcolor{yellow}0.017 & \cellcolor{yellow}0.013 & \cellcolor{yellow}0.013 & \cellcolor{yellow}0.153 & \cellcolor{yellow}0.022 & \cellcolor{yellow}0.016 & \cellcolor{yellow}0.013 \\ 
	& &	&  \cellcolor{skyblue}Mean & \cellcolor{skyblue}0.067 & \cellcolor{skyblue}0.014 & \cellcolor{skyblue}0.033 & \cellcolor{skyblue}0.024 & \cellcolor{skyblue}\cellcolor{skyblue}0.153 & \cellcolor{skyblue}0.022 & \cellcolor{skyblue}0.016 & \cellcolor{skyblue}0.016 \\ 
	& &	&  3rd.Qu & 0.076 & 0.021 & 0.017 & 0.017 & 0.189 & 0.028 & 0.022 & 0.023 \\ 
	& &	&  Max & 0.118 & 0.035 & 0.405 & 0.267 & 0.224 & 0.035 & 0.040 & 0.038 \\ 
\cmidrule{3-12}
	& & $\overline{S}_{0.05}$ &  Min & 1.055 & 1.200 & 1.261 & 1.329 & 1.128 & 1.196 & 1.256 & 1.317 \\ 
	& & &  1st.Qu & 1.142 & 1.479 & 1.624 & 1.680 & 1.313 & 1.614 & 1.847 & 1.811 \\ 
	& & &  \cellcolor{yellow}Median & \cellcolor{yellow}1.405 & \cellcolor{yellow}1.850 & \cellcolor{yellow}2.198 & \cellcolor{yellow}2.299 & \cellcolor{yellow}1.638 & \cellcolor{yellow}2.184 & \cellcolor{yellow}2.905 & \cellcolor{yellow}3.044 \\ 
	& & &  \cellcolor{skyblue}Mean & \cellcolor{skyblue}2.866 & \cellcolor{skyblue}3.213 & \cellcolor{skyblue}2.800 & \cellcolor{skyblue}3.583 & \cellcolor{skyblue}2.868 & \cellcolor{skyblue}3.273 & \cellcolor{skyblue}3.936 & \cellcolor{skyblue}4.571 \\ 
	& & &  3rd.Qu & 1.687 & 2.359 & 3.053 & 3.352 & 1.894 & 3.135 & 3.808 & 4.401 \\ 
	& & &  Max & 17.233 & 20.137 & 7.629 & 20.890 & 19.716 & 16.495 & 17.407 & 17.261 \\ 
\bottomrule
\end{longtable}
\end{small}
\end{center}

\vspace{-.4in}

In Table~\ref{tab:tune_2}, we present the selected tuning parameters $\xi_{\alpha,h}$ for different forecast horizons. Again, we report $h=1,5,10,15,20$. As $h$ increases, the selected tuning parameters increase to reflect greater uncertainty in the point forecast. In comparison with the tuning parameters in Table~\ref{tab:tune_1}, the following values are generally larger since the absolute residuals tend to be greater for an under-fitted model.
\begin{center}
\tabcolsep 0.13in
\renewcommand{\arraystretch}{1.05}
\begin{small}
\begin{longtable}{@{}llllrrrrrrrr}
\caption{\small At the nominal coverage probabilities of 80\% and 95\%, we present the selected tuning parameter $\xi_{\alpha,h}$ at $h=1,5,10,15$ and $20$. Based on the summary statistics, we present results from the functional time-series forecasting method, with $K$ determined via the EVR criterion.}\label{tab:tune_2}\\
  \toprule
&  	&	  & \multicolumn{4}{c}{Summary statistics (Female)} & \multicolumn{4}{c}{Summary statistics (Male)} \\
	\cmidrule(lr){4-7}\cmidrule(lr){8-11}
$K$ & $\alpha$ & $h$ & Quant & sd  & IQR & MAD  & Quant & sd  & IQR & MAD \\
  \midrule
EVR &  0.2 &  1 & 1.000 & 3.413 & 2.706 & 3.697 & 1.000 & 4.006 & 3.152 & 4.188 \\ 
   	&	& 5 & 1.012 & 4.038 & 3.334 & 4.275 & 1.010 & 5.513 & 4.500 & 5.713 \\ 
   	&	& 10 & 1.009 & 6.022 & 4.788 & 6.400 & 1.011 & 7.950 & 5.813 & 8.225 \\ 
   	&	& 15 & 1.016 & 8.025 & 6.850 & 9.000 & 1.018 & 10.788 & 8.550 & 13.400 \\ 
   	&	& 20 & 1.050 & 22.900 & 9.089 & 9.089 & 1.050 & 28.100 & 39.300 & 26.500 \\ 
\cmidrule{2-11}
   	&0.05 & 1 & 1.000 & 4.725 & 3.950 & 5.426 & 1.000 & 5.375 & 4.325 & 6.113 \\ 
  	& 	& 5 & 1.015 & 5.475 & 4.875 & 6.800 & 1.019 & 7.225 & 6.550 & 8.150 \\ 
  	&  	&10 & 1.035 & 8.400 & 7.913 & 10.550 & 1.025 & 11.300 & 12.100 & 16.800 \\ 
  	&  	&15 & 1.041 & 10.350 & 10.800 & 17.100 & 1.036 & 16.850 & 17.500 & 27.700 \\ 
  	&  	&20 & 20.802 & 103.300 & 9.089 & 9.089 & 20.802 & 103.300 & 154.500 & 103.300 \\ 
\bottomrule
\end{longtable}
\end{small}
\end{center}

\vspace{-.4in}

\subsection{Comparison between the number of tuning parameters}

Using the expanding-window forecast scheme, we evaluate the interval forecast accuracy, obtained from the functional time-series forecasting method with $K=6$, between the selections of a single tuning parameter $\xi_{\alpha, h}$ and double tuning parameters ($\underline{\xi}_{\alpha, h}, \overline{\xi}_{\alpha, h})$. From Table~\ref{tab:4}, the results remain similar. Thus, it is appropriate and computationally simpler to optimize a single tuning parameter.
\begin{center}
\renewcommand{\arraystretch}{1.1}
\begin{small}
\tabcolsep 0.08in
\begin{longtable}{@{}llllrrrrrrrr}
\caption{\small At the nominal coverage probabilities of 80\% and 95\%, we compare the interval forecast accuracy, as measured by the ECP, CPD, and interval scores over the 20 forecast horizons. Based on the summary statistics, we compare the difference between the selections of single and double tuning parameters.}\label{tab:4}\\
  \toprule
Tuning & &  		&  & \multicolumn{4}{c}{Summary statistics (Female)} & \multicolumn{4}{c}{Summary statistics (Male)} \\
	\cmidrule(lr){5-8}\cmidrule(lr){9-12}
Parameter & $\alpha$ & Metric & Summary & Quant & sd  & IQR & MAD  & Quant & sd  & IQR & MAD \\
  \midrule
Double & 0.2 & ECP & Min & 0.693 & 0.748 & 0.789 & 0.777 & 0.743 & 0.158 & 0.803 & 0.158 \\ 
	& & &  1st.Qu & 0.783 & 0.810 & 0.813 & 0.815 & 0.784 & 0.827 & 0.835 & 0.834 \\ 
	& & &  \cellcolor{yellow}Median & \cellcolor{yellow}0.792 & \cellcolor{yellow}0.847 & \cellcolor{yellow}0.827 & \cellcolor{yellow}0.850 & \cellcolor{yellow}0.798 & \cellcolor{yellow}0.865 & \cellcolor{yellow}0.852 & \cellcolor{yellow}0.857 \\ 
	& & &  \cellcolor{skyblue}Mean & \cellcolor{skyblue}0.787 & \cellcolor{skyblue}0.837 & \cellcolor{skyblue}0.838 & \cellcolor{skyblue}0.846 & \cellcolor{skyblue}0.796 & \cellcolor{skyblue}0.796 & \cellcolor{skyblue}0.851 & \cellcolor{skyblue}0.819 \\ 
	& & &  3rd.Qu. & 0.801 & 0.872 & 0.861 & 0.865 & 0.813 & 0.890 & 0.866 & 0.871 \\ 
	& & &  Max & 0.812 & 0.891 & 0.921 & 0.893 & 0.833 & 0.911 & 0.902 & 0.897 \\ 
\cmidrule{3-12}
	& & CPD & Min. & 0.000 & 0.001 & 0.000 & 0.006 & 0.000 & 0.000 & 0.003 & 0.001 \\ 
	& & & 1st.Qu & 0.006 & 0.027 & 0.013 & 0.016 & 0.007 & 0.043 & 0.035 & 0.046 \\ 
	& & &  \cellcolor{yellow}Median & \cellcolor{yellow}0.012 & \cellcolor{yellow}0.049 & \cellcolor{yellow}0.027 & \cellcolor{yellow}0.050 & \cellcolor{yellow}0.013 & \cellcolor{yellow}0.084 & \cellcolor{yellow}0.052 & \cellcolor{yellow}0.061 \\ 
	& & &  \cellcolor{skyblue}Mean & \cellcolor{skyblue}0.016 & \cellcolor{skyblue}0.049 & \cellcolor{skyblue}0.039 & \cellcolor{skyblue}0.048 & \cellcolor{skyblue}0.019 & \cellcolor{skyblue}0.121 & \cellcolor{skyblue}0.051 & \cellcolor{skyblue}0.085 \\ 
	& & &  3rd.Qu & 0.017 & 0.072 & 0.061 & 0.065 & 0.028 & 0.095 & 0.066 & 0.073 \\ 
	& & &  Max & 0.107 & 0.091 & 0.121 & 0.093 & 0.057 & 0.642 & 0.102 & 0.642 \\   
\cmidrule{3-12}
	& & $\overline{S}_{0.2}$ & Min & 0.500 & 0.493 & 0.498 & 0.502 & 0.406 & 0.410 & 0.422 & 0.415 \\ 
	& & &  1st.Qu & 0.576 & 0.597 & 0.591 & 0.609 & 0.569 & 0.559 & 0.591 & 0.606 \\ 
	& & &  \cellcolor{yellow}Median & \cellcolor{yellow}0.724 & \cellcolor{yellow}0.749 & \cellcolor{yellow}0.782 & \cellcolor{yellow}0.817 & \cellcolor{yellow}0.770 & \cellcolor{yellow}0.761 & \cellcolor{yellow}0.775 & \cellcolor{yellow}0.820 \\ 
	& & &  \cellcolor{skyblue}Mean & \cellcolor{skyblue}0.747 & \cellcolor{skyblue}0.856 & \cellcolor{skyblue}0.901 & \cellcolor{skyblue}1.071 & \cellcolor{skyblue}0.774 & \cellcolor{skyblue}0.812 & \cellcolor{skyblue}1.026 & \cellcolor{skyblue}1.112 \\ 
	& & &  3rd.Qu & 0.883 & 0.943 & 0.979 & 1.047 & 0.983 & 0.871 & 0.959 & 1.024 \\ 
	& & &  Max & 1.181 & 2.403 & 2.338 & 4.457 & 1.141 & 1.848 & 4.052 & 5.742 \\   	
\midrule
	& 0.05 & ECP &  Min & 0.693 & 0.931 & 0.933 & 0.931 & 0.792 & 0.158 & 0.920 & 0.158 \\ 
	& & &  1st.Qu & 0.868 & 0.948 & 0.944 & 0.942 & 0.876 & 0.929 & 0.958 & 0.955 \\ 
	& & &  \cellcolor{yellow}Median & \cellcolor{yellow}0.900 & \cellcolor{yellow}0.958 & \cellcolor{yellow}0.952 & \cellcolor{yellow}0.951 & \cellcolor{yellow}0.907 & \cellcolor{yellow}0.978 & \cellcolor{yellow}0.975 & \cellcolor{yellow}0.974 \\ 
	& & &  \cellcolor{skyblue}Mean & \cellcolor{skyblue}0.882 & \cellcolor{skyblue}0.958 & \cellcolor{skyblue}0.954 & \cellcolor{skyblue}0.954 & \cellcolor{skyblue}0.893 & \cellcolor{skyblue}0.889 & \cellcolor{skyblue}0.969 & \cellcolor{skyblue}0.930 \\ 
	& & &  3rd.Qu & 0.910 & 0.967 & 0.962 & 0.967 & 0.920 & 0.992 & 0.989 & 0.987 \\ 
	& & &  Max & 0.916 & 0.984 & 0.980 & 0.983 & 0.936 & 0.999 & 0.997 & 0.998 \\ 
\cmidrule{3-12}
	& & CPD & Min & 0.034 & 0.000 & 0.000 & 0.000 & 0.014 & 0.001 & 0.004 & 0.012 \\ 
	& & &  1st.Qu & 0.040 & 0.006 & 0.005 & 0.007 & 0.030 & 0.025 & 0.019 & 0.018 \\ 
	& & &  \cellcolor{yellow}Median & \cellcolor{yellow}0.050 & \cellcolor{yellow}0.012 & \cellcolor{yellow}0.010 & \cellcolor{yellow}0.009 & \cellcolor{yellow}0.043 & \cellcolor{yellow}0.032 & \cellcolor{yellow}0.026 & \cellcolor{yellow}0.030 \\ 
	& & &  \cellcolor{skyblue}Mean & \cellcolor{skyblue}0.068 & \cellcolor{skyblue}0.013 & \cellcolor{skyblue}0.012 & \cellcolor{skyblue}0.012 & \cellcolor{skyblue}0.057 & \cellcolor{skyblue}0.105 & \cellcolor{skyblue}0.027 & \cellcolor{skyblue}0.066 \\ 
	& & &  3rd.Qu & 0.082 & 0.020 & 0.017 & 0.018 & 0.074 & 0.047 & 0.039 & 0.038 \\ 
	& & &  Max & 0.257 & 0.034 & 0.030 & 0.033 & 0.158 & 0.792 & 0.047 & 0.792 \\ 
\cmidrule{3-12}
	& & $\overline{S}_{0.05}$ & Min & 0.864 & 0.793 & 0.773 & 0.800 & 0.625 & 0.619 & 0.679 & 0.652 \\ 
	& & &  1st.Qu & 0.921 & 0.991 & 0.975 & 1.051 & 0.798 & 1.076 & 0.966 & 1.132 \\ 
	& & &  \cellcolor{yellow}Median & \cellcolor{yellow}1.138 & \cellcolor{yellow}1.218 & \cellcolor{yellow}1.317 & \cellcolor{yellow}1.530 & \cellcolor{yellow}1.066 & \cellcolor{yellow}1.251 & \cellcolor{yellow}1.555 & \cellcolor{yellow}1.748 \\ 
	& & &  \cellcolor{skyblue}Mean & \cellcolor{skyblue}1.274 & \cellcolor{skyblue}1.740 & \cellcolor{skyblue}1.832 & \cellcolor{skyblue}2.606 & \cellcolor{skyblue}1.093 & \cellcolor{skyblue}2.038 & \cellcolor{skyblue}2.509 & \cellcolor{skyblue}3.180 \\ 
	& & &  3rd.Qu & 1.416 & 1.805 & 1.896 & 2.076 & 1.316 & 2.023 & 2.119 & 3.034 \\ 
	& & &  Max & 2.808 & 7.591 & 8.413 & 15.355 & 1.795 & 6.878 & 15.523 & 21.159 \\ 	
\bottomrule
\end{longtable}
\end{small}
\end{center}

\vspace{-.4in}

In Table~\ref{tab:tune_3}, we present the selected tuning parameters $(\underline{\xi}_{\alpha, h},\overline{\xi}_{\alpha, h})$ for different forecast horizons. Again, we report $h=1,5,10,15,20$. When the summary statistics are sd, IQR, and MAD, as $h$ increases, the selected tuning parameters generally increase to reflect greater uncertainty in the point forecast. When the summary statistics are quantile, the selected tuning parameters do not increase much as $h$ increases. This is because quantile is a more adaptive summary statistic than the others.
\begin{center}
\tabcolsep 0.11in
\renewcommand{\arraystretch}{1.2}
\begin{small}
\begin{longtable}{@{}lllllrrrrrrrr}
\caption{\small At the nominal coverage probabilities of 80\% and 95\%, we present the selected tuning parameters $(\underline{\xi}_{\alpha,h}, \overline{\xi}_{\alpha,h})$ at $h=1,5,10,15$ and $20$. Based on the summary statistics, we present the results using the functional time-series forecasting method with $K=6$.}\label{tab:tune_3}\\
  \toprule
Tuning  & & & & \multicolumn{2}{c}{Quant} & \multicolumn{2}{c}{sd}  & \multicolumn{2}{c}{IQR} & \multicolumn{2}{c}{MAD} \\
\cmidrule(lr){5-6}\cmidrule(lr){7-8}\cmidrule(lr){9-10}\cmidrule(lr){11-12}
Parameter & $\alpha$ & Sex &  $h$ & $\underline{\xi}_{\alpha,h}$ & $\overline{\xi}_{\alpha,h}$ & $\underline{\xi}_{\alpha,h}$ & $\overline{\xi}_{\alpha,h}$ & $\underline{\xi}_{\alpha,h}$ & $\overline{\xi}_{\alpha,h}$ & $\underline{\xi}_{\alpha,h}$ & $\overline{\xi}_{\alpha,h}$ \\
\midrule
Double 	& 0.2 & F & 1 & 0.999 & 1.044 & 1.336 & 1.361 & 1.326 & 0.939 & 1.244 & 1.682 \\ 
		& 	& 	&   5 & 1.071 & 1.000 & 1.131 & 2.144 & 1.106 & 1.469 & 1.628 & 1.731 \\ 
		& 	& 	&   10 & 1.070 & 1.007 & 1.131 & 3.756 & 2.067 & 1.596 & 1.359 & 4.029 \\ 
		& 	& 	&   15 & 1.044 & 1.048 & 1.418 & 2.855 & 1.817 & 1.958 & 2.695 & 2.549 \\ 
		& 	& 	&  20 & 0.925 & 1.225 & 7.326 & 14.105 & 13.566 & 15.234 & 9.535 & 10.989 \\ 
\cmidrule{3-12}
		& 	& M  & 1 & 1.012 & 0.991 & 1.624 & 1.210 & 1.240 & 1.068 & 1.731 & 1.172 \\ 
		& 	&   	& 5 & 1.003 & 1.092 & 2.396 & 1.273 & 2.057 & 1.317 & 2.976 & 1.511 \\ 
		& 	&   	& 10 & 1.036 & 0.938 & 2.352 & 2.939 & 2.062 & 1.118 & 3.764 & 1.487 \\ 
		& 	&   	& 15 & 1.015 & 1.049 & 2.836 & 1.595 & 1.992 & 1.221 & 3.752 & 1.481 \\ 
		& 	&   	& 20 & 1.103 & 0.941 & 1.250 & 1.050 & 50.178 & 22.379 & 1.200 & 1.000 \\ 
\cmidrule{2-12}
& 0.05 	&   F & 1 	& 1.000 & 1.000 & 2.059 & 2.115 	& 1.702 & 1.971 & 2.175 & 2.848 \\ 
		& 	&  	&  5 	& 1.025 & 1.025 & 1.924 & 3.683 	& 1.711 & 3.078 & 2.588 & 3.069 \\ 
		& 	&   	& 10 & 1.006 & 1.131 & 2.054 & 6.464 	& 3.839 & 2.478 & 2.605 & 7.395 \\ 
		& 	&  	& 15 & 1.061 & 1.092 & 2.149 & 6.822 	& 3.025 & 3.575 & 6.092 & 4.824 \\ 
		& 	&  	& 20 & 1.037 & 1.175 & 24.320 & 48.862 	& 52.585 & 61.528 & 35.266 & 41.359 \\ 
\cmidrule{3-12}
		&  	& M  & 1 & 1.000 & 1.000 & 2.847 & 1.541 & 2.573 & 1.282 & 3.252 & 1.705 \\ 
		& 	&   	& 5 & 1.019 & 1.047 & 6.123 & 1.684 & 4.664 & 1.586 & 7.558 & 2.065 \\ 
		& 	&   	& 10 & 1.062 & 0.962 & 3.308 & 4.060 & 4.249 & 2.073 & 11.548 & 2.729 \\ 
		& 	&  	& 15 & 1.132 & 0.985 & 8.479 & 2.807 & 4.531 & 2.294 & 12.810 & 3.125 \\ 
		& 	&   	& 20 & 1.150 & 1.025 & 1.250 & 1.050 & 202.631 & 86.177 & 1.200 & 1.000 \\ 
\bottomrule
\end{longtable}
\end{small}
\end{center}

\vspace{-.4in}

\subsection{Comparison between the split and sequential conformal predictions}

In Table~\ref{tab:5}, using the expanding-window forecast scheme, we evaluate the interval forecast accuracy obtained from the functional time-series forecasting method with $K=6$ between the split and sequential conformal predictions. Compared with split conformal prediction, sequential conformal prediction achieves higher interval forecast accuracy. At the nominal coverage probabilities of 80\% and 95\%, the empirical coverage probabilities from the sequential conformal prediction are not below the nominal one, while the interval scores are \textit{smaller} and \textit{more stable} than the ones from the split conformal prediction. 
\begin{table}[!htb]
\centering
\tabcolsep 0.13in
\caption{\small At the nominal coverage probabilities of 80\% and 95\%, we compare the interval forecast accuracy, as measured by the ECP, CPD, and mean interval score, over the 20 forecast horizons. Based on the summary statistics, we compare the difference between the split and sequential conformal predictions.}\label{tab:5}
\begin{small}
\renewcommand{\arraystretch}{1.9}
\begin{tabular}{@{}lllrrrrrrrr}
\toprule
& & &  \multicolumn{4}{c}{Summary statistics (Female)} & \multicolumn{4}{c}{Summary statistics (Male)} \\
\cmidrule(lr){4-7}\cmidrule(lr){8-11}
Statistics & $\alpha$ & Summary & ECP & CPD & width & score & ECP & CPD & width & score \\ 
\midrule
Quantile 	& 0.2 	& Min 	& 0.795 & 0.038 & 0.292 & 0.513 & 0.808 & 0.028 & 0.274 & 0.424 \\ 
		& 		& 1st.Qu 	& 0.833 & 0.047 & 0.389 & 0.582 & 0.823 & 0.050 & 0.417 & 0.540 \\ 
		&		&  \cellcolor{yellow}Median & \cellcolor{yellow}0.855 & \cellcolor{yellow}0.055 & \cellcolor{yellow}0.527 & \cellcolor{yellow}0.718 & \cellcolor{yellow}0.827 & \cellcolor{yellow}0.067 & \cellcolor{yellow}0.587 & \cellcolor{yellow}0.727 \\ 
		&		&  \cellcolor{skyblue}Mean & \cellcolor{skyblue}0.850 & \cellcolor{skyblue}0.058 & \cellcolor{skyblue}0.549 & \cellcolor{skyblue}0.725 & \cellcolor{skyblue}0.830 & \cellcolor{skyblue}0.066 & \cellcolor{skyblue}0.594 & \cellcolor{skyblue}0.732 \\ 
		&		&  3rd.Qu 	& 0.862 & 0.062 & 0.698 & 0.863 & 0.840 & 0.082 & 0.763 & 0.902 \\ 
		&		&  Max 	& 0.908 & 0.108 & 0.862 & 0.982 & 0.851 & 0.103 & 0.929 & 1.062 \\ 
\\
		& 0.05 	& Min 	& 0.938 & 0.014 & 0.465 & 0.855 & 0.901 & 0.009 & 0.456 & 0.627 \\ 
		&		& 1st.Qu 	& 0.950 & 0.016 & 0.598 & 0.897 & 0.939 & 0.030 & 0.679 & 0.798 \\ 
		& 		&  \cellcolor{yellow}Median & \cellcolor{yellow}0.956 & \cellcolor{yellow}0.020 & \cellcolor{yellow}0.766 & \cellcolor{yellow}1.082 & \cellcolor{yellow}0.948 & \cellcolor{yellow}0.034 & \cellcolor{yellow}0.878 & \cellcolor{yellow}1.026 \\ 
		&		&  \cellcolor{skyblue}Mean & \cellcolor{skyblue}0.959 & \cellcolor{skyblue}0.022 & \cellcolor{skyblue}0.792 & \cellcolor{skyblue}1.075 & \cellcolor{skyblue}0.942 & \cellcolor{skyblue}0.034 & \cellcolor{skyblue}0.900 & \cellcolor{skyblue}1.062 \\ 
		&		& 3rd.Qu 	& 0.967 & 0.022 & 0.944 & 1.189 & 0.954 & 0.036 & 1.127 & 1.295 \\ 
		&		& Max 	& 0.990 & 0.040 & 1.233 & 1.427 & 0.961 & 0.049 & 1.345 & 1.530 \\ 
\bottomrule
\end{tabular}
\end{small}
\end{table}
  
\begin{figure}[!htb]
\centering
\includegraphics[width=8.75cm]{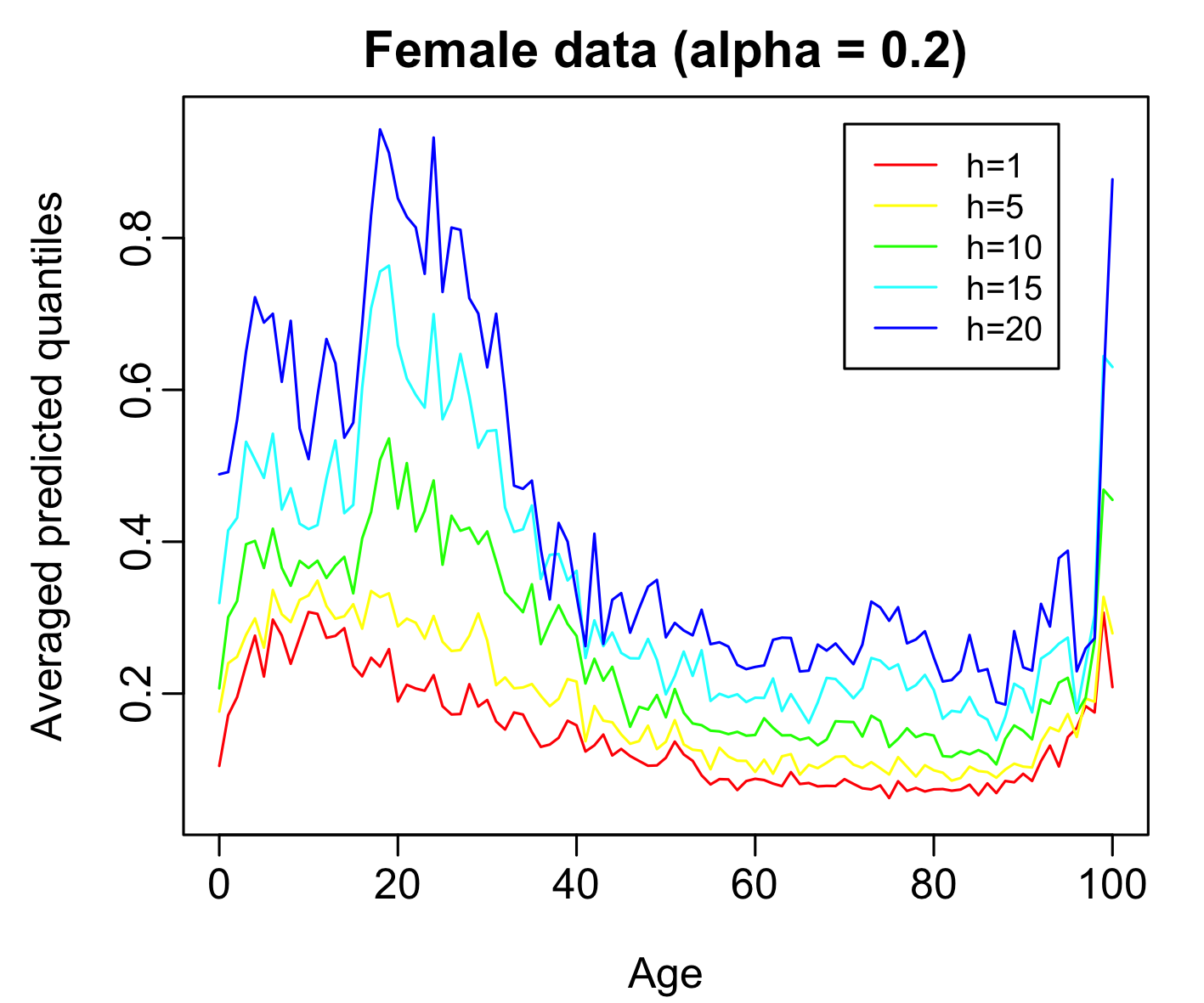}
\quad
\includegraphics[width=8.75cm]{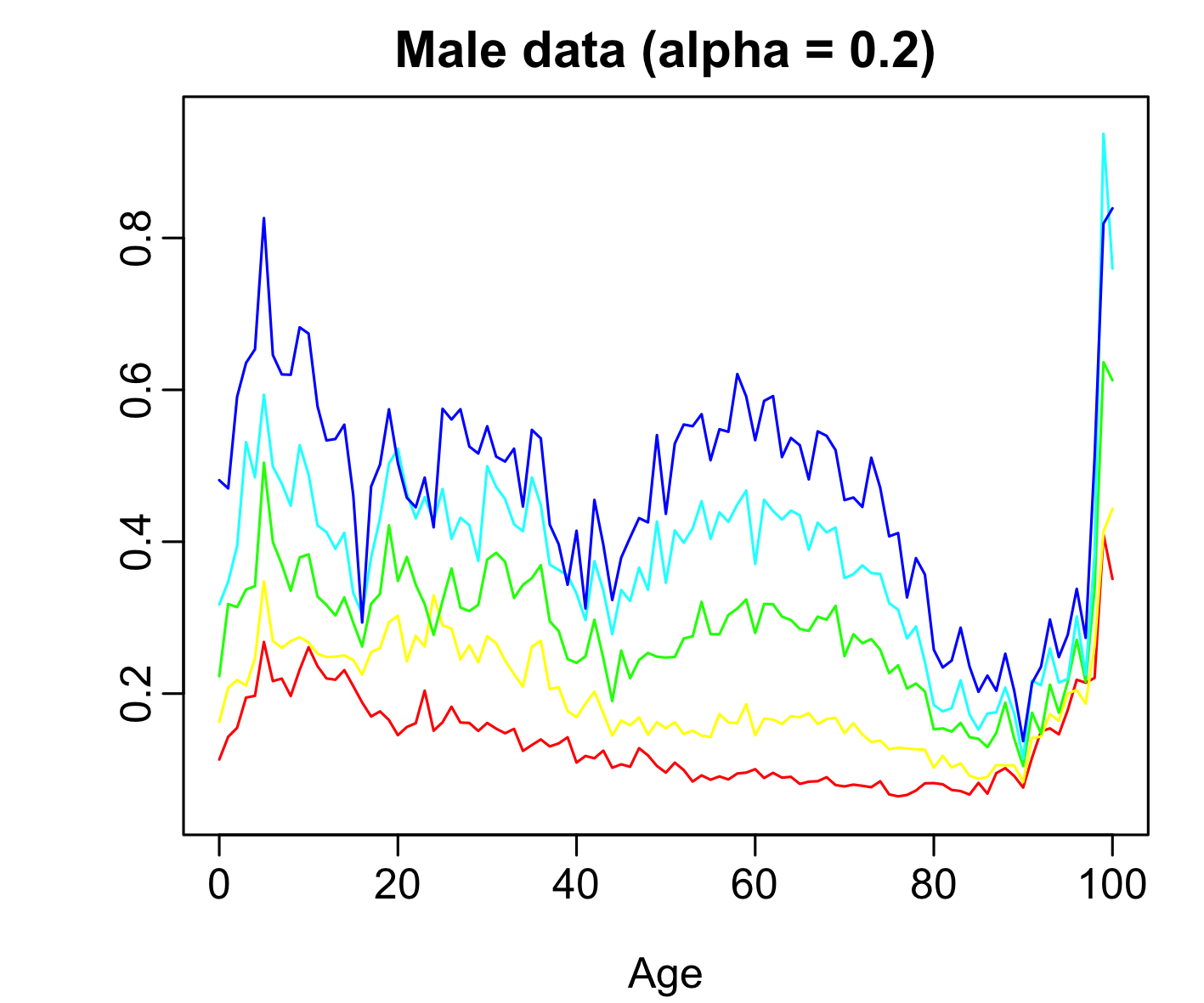}
\\
\includegraphics[width=8.75cm]{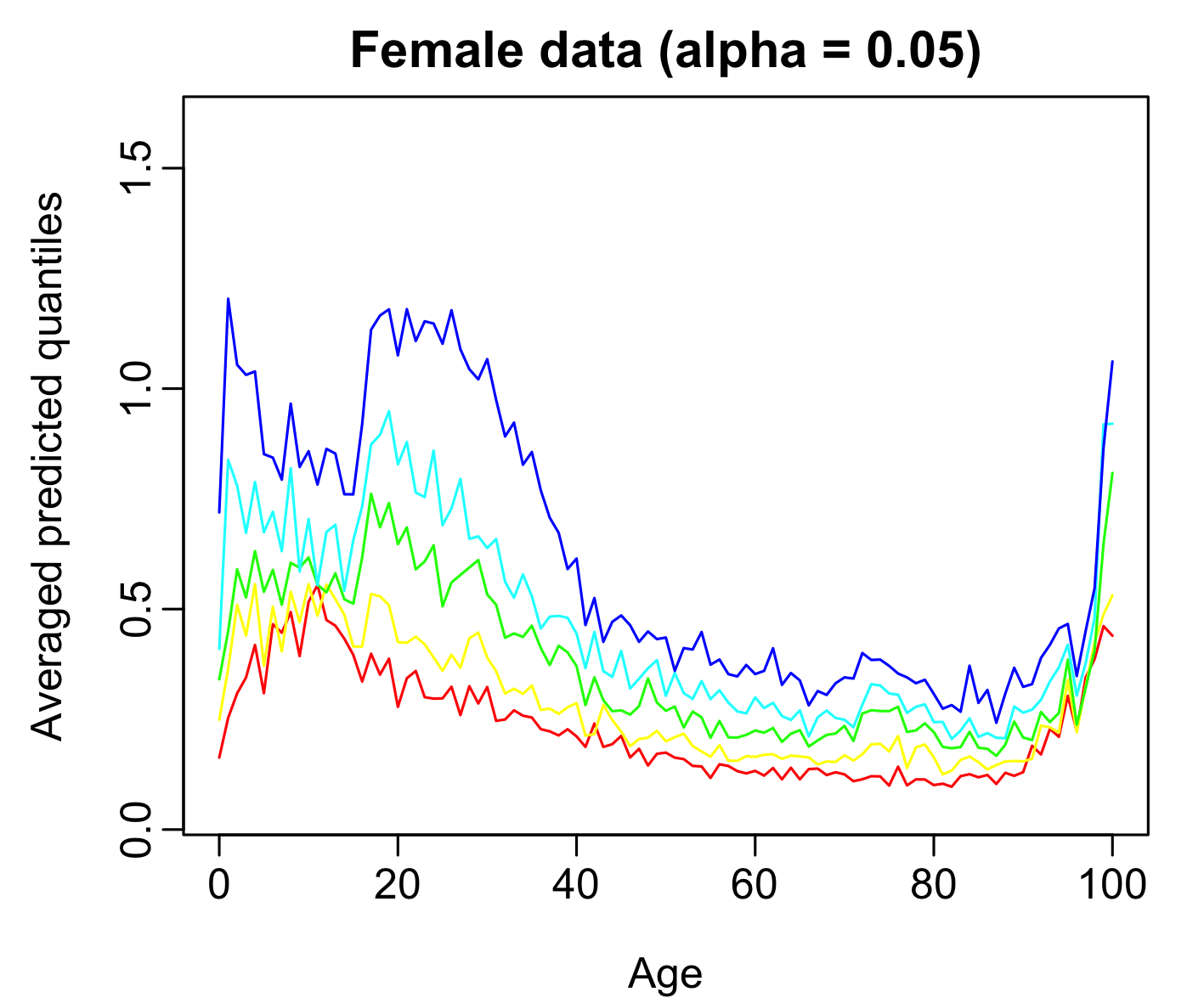}
\quad
\includegraphics[width=8.75cm]{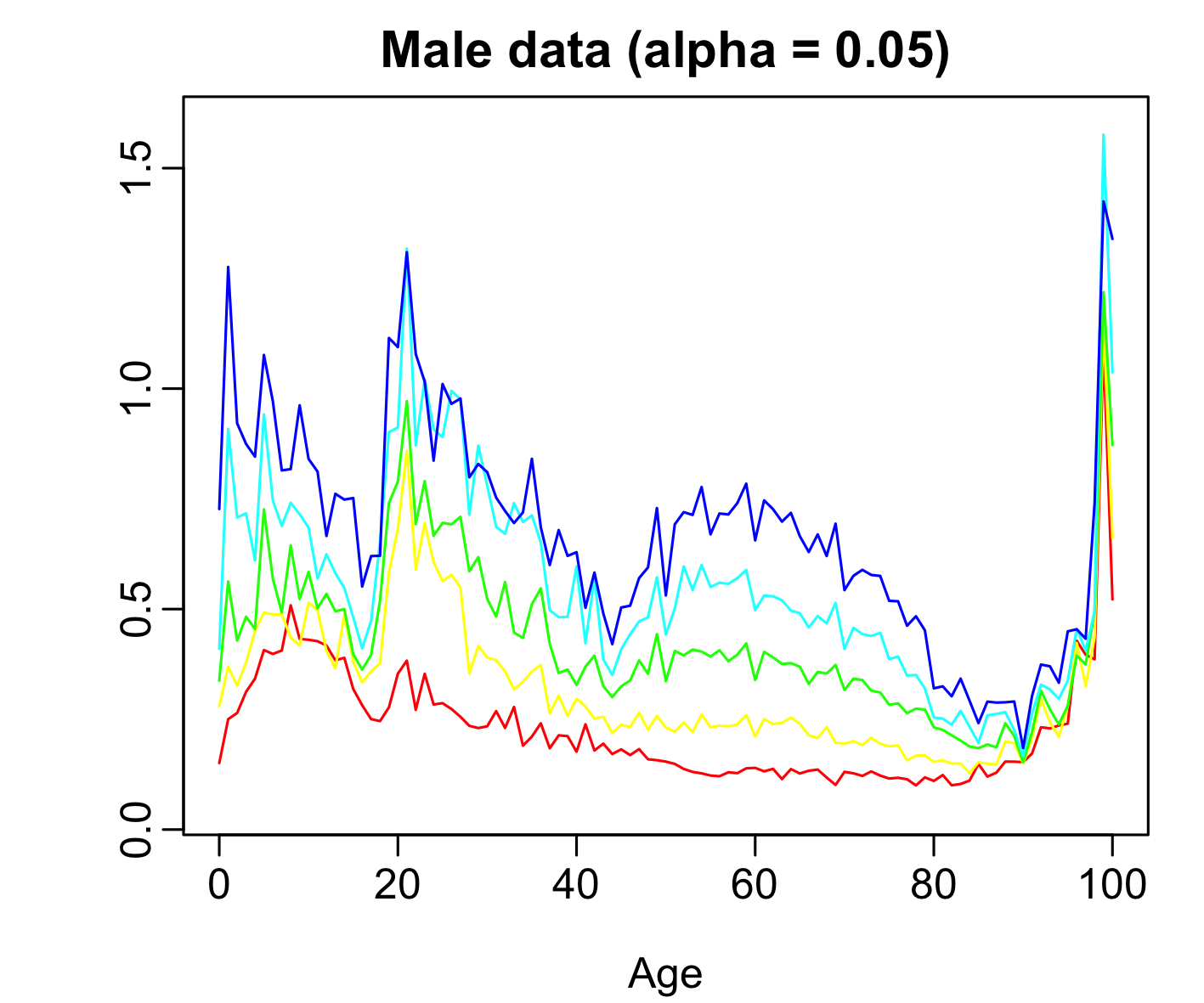}
\caption{\small Under the nominal coverage probabilities of 80\% and 95\%, we present the age-specific predicted quantiles averaged over the number of testing data for $h=1,5,10,15$ and 20.}\label{fig:4}
\end{figure}

In Figure~\ref{fig:4}, we present the averaged predicted quantiles over the number of testing data for $h=1,5,10,15$ and 20 under the nominal coverage probabilities of 80\% and 95\%. In general, the predicted quantiles are larger for ages between 0 and 50 than those for ages above 40, except for the highest ages. Between the two levels of significance, the magnitude differs, but the patterns are generally similar.
  
\section{Conclusion}\label{sec:5}

We present conformal prediction as a means to quantify forecast uncertainty for functional time series and demonstrate its use using Australian age- and sex-specific log mortality rates. The conformal prediction framework is model-agnostic and distribution-free to construct prediction intervals. Using the functional time-series method as a forecasting model, we compare the finite-sample performance of split and sequential conformal predictions. 

As part of the sensitivity analyses of the split conformal prediction, we examine the interval forecast accuracy between the expanding- and rolling-window forecast schemes and find a marginal difference. When the number of components is selected by the EVR criterion, it can lead to under-fitting, an issue in the male data when quantiles are used as summary statistics. Between the selections of one and two tuning parameters, there is a marginal difference; thus, it is computationally faster to calibrate a single tuning parameter in the validation set.

The split conformal prediction requires sample splitting, which can lead to inferior interval forecast accuracy at longer forecast horizons. As a model-agnostic and tuning parameter-free approach, the sequential conformal prediction gradually updates the predictive quantiles as new data sequentially arrive. Because it does not require calibration using the validation set, it is our recommended approach for quantifying finite-sample prediction uncertainty.

There are various ways in which the current paper can be further extended, and we briefly outline three:
\begin{inparaenum}[1)]
\item We demonstrate conformal prediction using Australian age- and sex-specific mortality rates at the national level; other datasets at the national and subnational levels are also possible.
\item We implement a univariate functional time-series forecasting method to model female and male data individually, but other \textit{multi-population} modeling techniques may also be considered.
\item In the sequential conformal prediction, we model the temporal dependence of the absolute residuals via an AR process in a quantile regression. Other time-series models can also be applied.
\end{inparaenum}

\section*{Acknowledgment}

The author is grateful to the comments provided by two anonymous reviewers, which led to a much-improved version. I thank the insightful comments received from the participants at the 2025 Australasian Actuarial Education and Research Symposium. This research is financially supported by the Australian Research Council Discovery Project DP230102250 and Future Fellowship FT240100338.

\newpage
\bibliographystyle{agsm}
\bibliography{CP_FTS.bib}

\end{document}